\documentclass[twocolumn]{revtex4}
\usepackage{amsmath}
\usepackage{amsfonts}
\usepackage{amssymb}
\usepackage{graphicx}
\graphicspath{{images/}} 
\usepackage{color}
\usepackage{url}
\usepackage{dsfont} 
\usepackage{xr}

\newcommand{\beq}{\begin{equation}}
\newcommand{\eeq}{\end{equation}}
\newcommand{\bea}{\begin{eqnarray}}
\newcommand{\eea}{\end{eqnarray}}

\DeclareMathOperator*{\argmax}{argmax}

\definecolor{red}{rgb}{1,0,0}

\renewcommand{\vec}[1]{{\boldsymbol{#1}}}
\renewcommand{\mathbb}[1]{\mathds{#1}}

\newcommand{\dsum}{\displaystyle\sum}
\newcommand{\dprod}{\displaystyle\prod}

\newcommand{\beginsupplement}{%
	\setcounter{table}{0}
	\renewcommand{\thetable}{S\arabic{table}}%
	\setcounter{figure}{0}
	\renewcommand{\thefigure}{S\arabic{figure}}%
}

\newcommand{\figone}{
\begin{figure*}[t]
	\noindent\includegraphics[width=\textwidth]{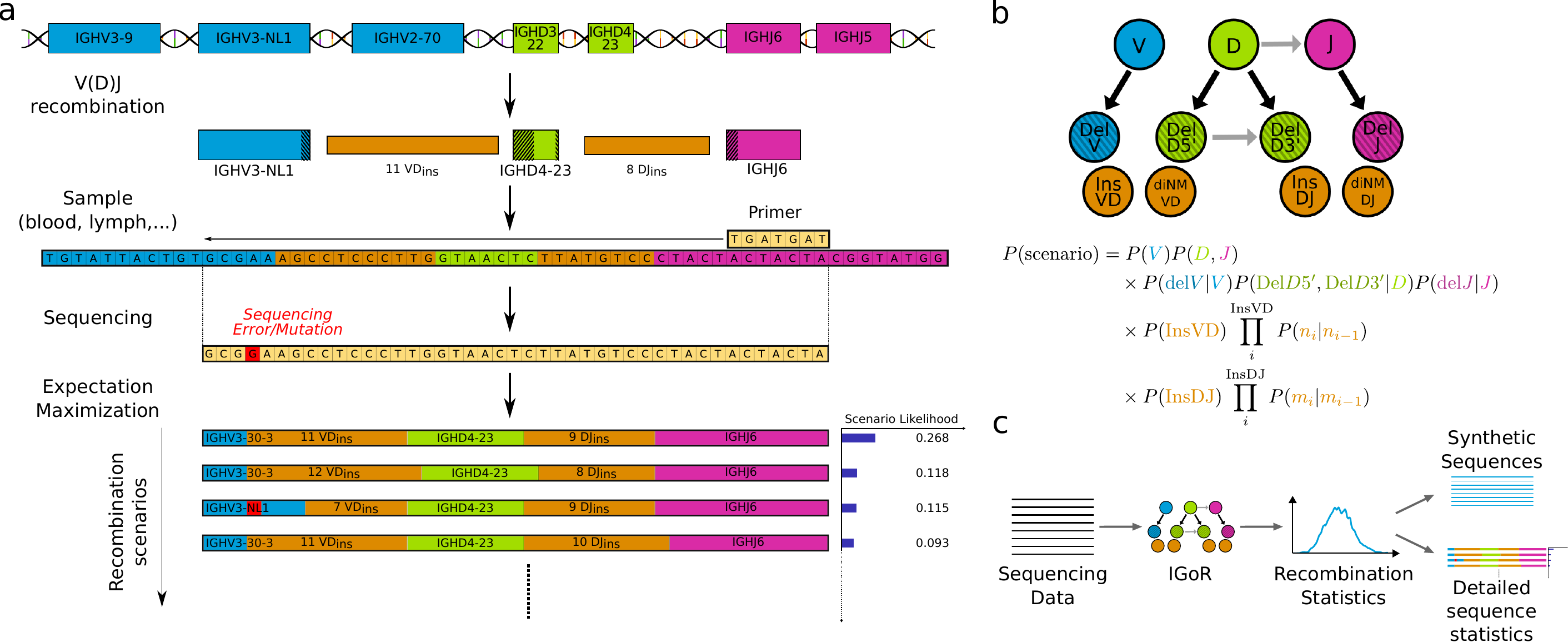}
	\caption{\textbf {IGoR's pipeline for sequence analysis.} \textbf{(a)} V(D)J recombination proceeds by joining randomly selected segments (V, D, and J segments in the case of IGH). Each segments gets trimmed at its ends (hashed areas), and a varying number of nontemplated insertions are added between them (orange). Hypermutations (in the case of B cells) or sequencing errors (in red) further enhance diversity.
IGoR lists putative recombination scenarios consistent with the observed sequence, and weighs them according to their likelihood.
\textbf{(b)} The likelihood of each scenario is computed using a Bayesian network of dependencies between the recombination features (V, D, J segment choices, insertions and deletions), as illustrated here for the human TRB locus. Architectures for TRA and IGH are described in Online Methods. {\bf (c)} IGoR's pipeline includes three modes. In the learning mode, IGoR learns recombination statistics from data sequences. In the analysis mode, IGoR outputs detailed recombination scenario statistics for each sequence. In the generation mode, IGoR produces synthetic sequences with specified recombination statistics.
		\label{Fig1}
	}
\end{figure*}
}

\newcommand{\figtwo}{
\begin{figure*}[t]
	\noindent\includegraphics[width=\textwidth]{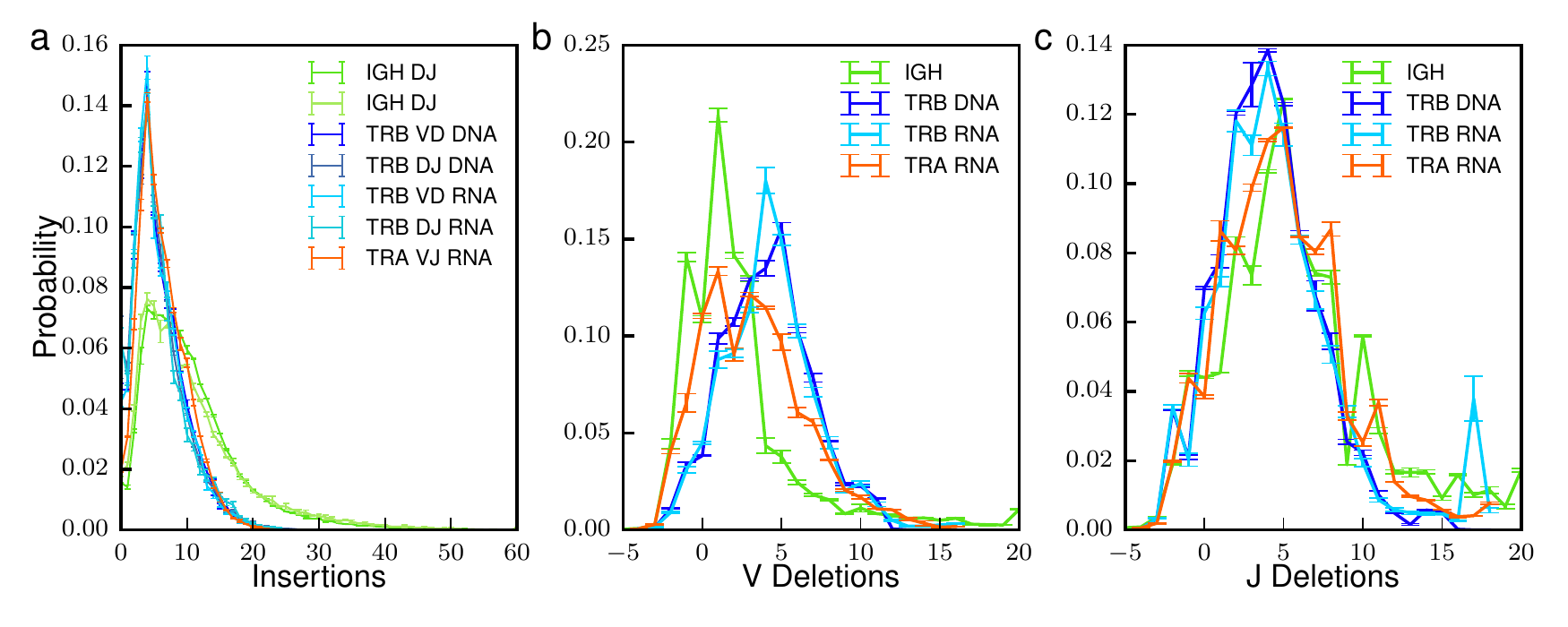}
	\caption{\textbf {IGoR infers reproducible recombination statistics.} \textbf{(a)} Distribution of the number of insertions at the junctions of recombined genes: IGH at the VD and DJ junctions from DNA data \cite{larimore_shaping_2012}, TRB at the VD and DJ junction from both DNA \cite{robins2009comprehensive} and mRNA data \cite{pogorelyy2016persisting}, and TRA at the VJ insertion site from mRNA data \cite{pogorelyy2016persisting}. \textbf{(b)},\textbf{(c)}. Average distribution of the number of deletions across {\bf (b)} V and {\bf (c)} J genes. Negative deletions correspond to palindromic insertions (P nucleotides), e.g. -2 means 2 P-nucleotides. The inferred distributions are robust to the choice of individuals, genetic material (mRNA or DNA) and sequencing technology. Error bars show 1 standard deviation across individuals.}
		\label{Fig2}
\end{figure*}
}

\newcommand{\figthree}{
\begin{figure}	\noindent\includegraphics[width=.5\textwidth]{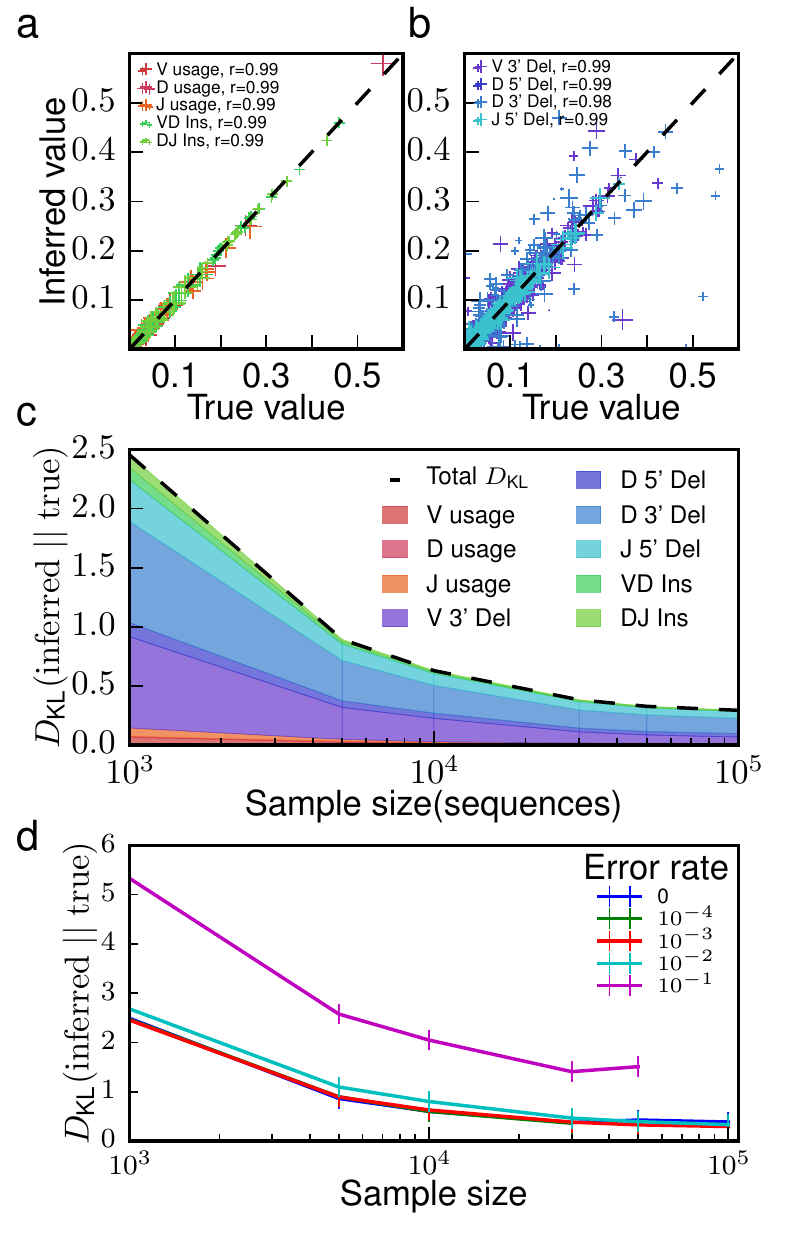}	
	\caption{\textbf{Validation on synthetic data.}
Short synthetic reads of recombined TRB or IGH sequences were generated with known recombination statistics, and given to IGoR as input to re-infer these statistics.  Inference with $10^5$ sequences and a typical sequencing error rate of $10^{-3}$ gives excellent agreement for  ({\bf a}) gene usage and insertion statistics and ({\bf b}) deletion statistics (Pearson's $r$ for deletions is calculated on the joint statistics of gene usage and deletion number; cross size scales with gene usage).  ({\bf c}) Discrepancy between true and inferred values of the recombination statistics, measured by the Kullback-Leibler divergence,  as a function of the number of unique sequences in the sample, and decomposed according to the features of the recombination scenario. ({\bf d}) Same as ({\bf c}), for increasing rates of sequencing errors or of hypermutations.
}
\label{Fig3}
\end{figure}
}

\newcommand{\figfour}{
\begin{figure*}
	\noindent\includegraphics[width=\textwidth]{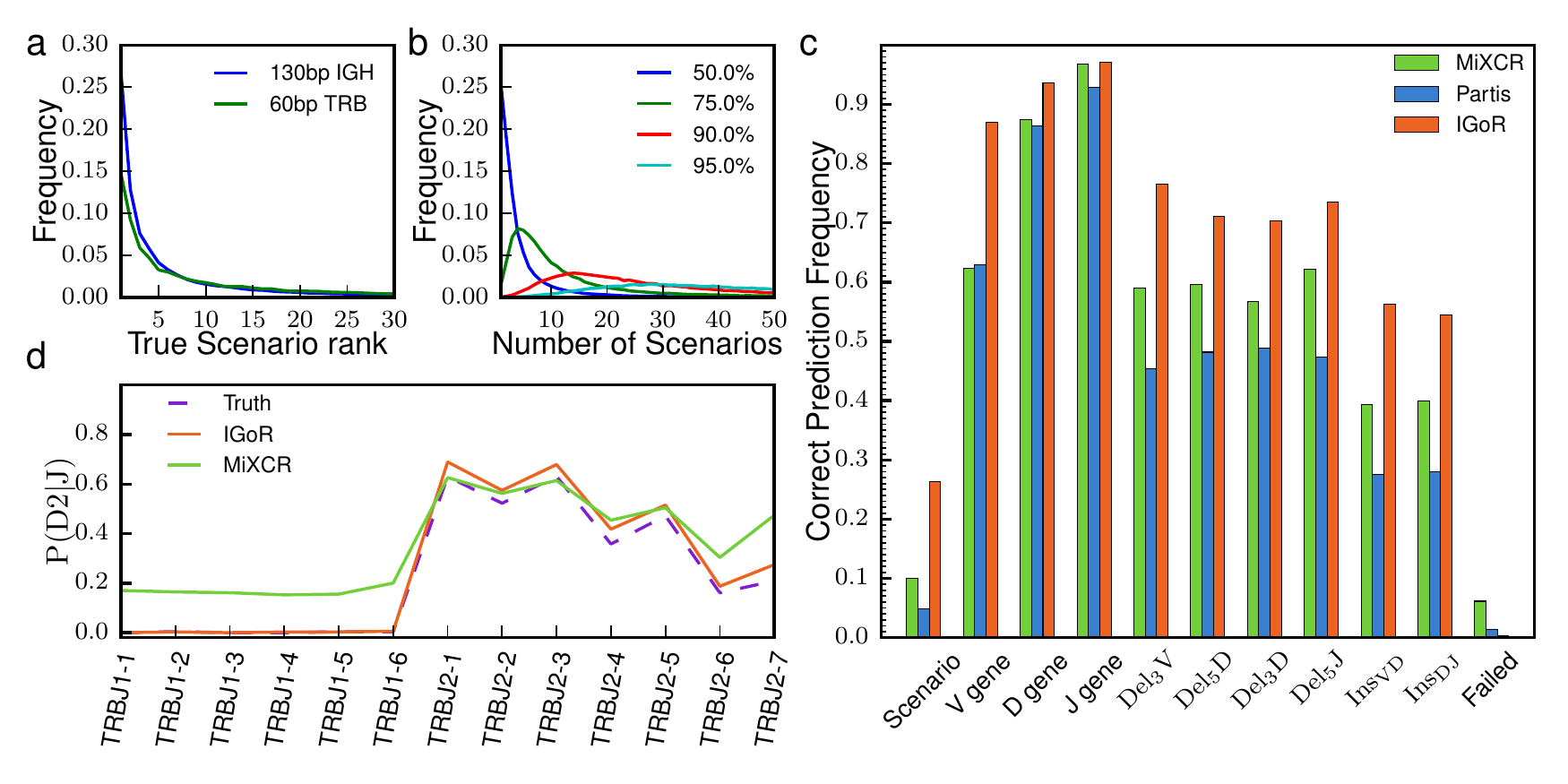}
	\caption{\textbf{Probabilistic analysis of putative recombination scenarios and comparison to existing methods.} 
Synthetic 130-bp reads of recombined IGH sequences and 60-bp reads of TRB sequences were generated with a $5\cdot 10^{-3}$ error rate, and processed for analysis by IGoR and two existing methods, MiXCR \cite{bolotin2015mixcr} and Partis \cite{ralph_consistency_2016}. IGoR ranks putative scenarios by descending order of likelihood.
(\textbf{a}) Distribution of the rank of the true scenario as called by IGoR. Note that the best-ranked (maximum-likelihood) scenario is the correct one in less than 30\% of cases.
		(\textbf{b}) Distribution of the number of scenarios that need to be enumerated (from most to least likely) to include the true scenario with 50\% (blue), 75\% (green), 90\% (red), or 95\% (cyan) confidence.
		(\textbf{c}) Frequency with which IGoR, MiXCR and Partis call the correct scenario of recombination as the most likely one (`scenario'), as well as each separate feature of the scenario (`V gene,' etc.).  `Failed' corresponds to sequences for which the algorithm did not output an assignment. 
		(\textbf{c}) Usage frequency of TRB D gene conditioned on the J gene, inferred by the IGoR and MiXCR (Partis does not handle TCR sequences). IGoR recovers the physiological exclusion between D2 and J1, while MiXCR does not.
	}
	\label{Fig4}
\end{figure*}
}

\newcommand{\figfive}{
\begin{figure*}
\noindent\includegraphics[width=\textwidth]{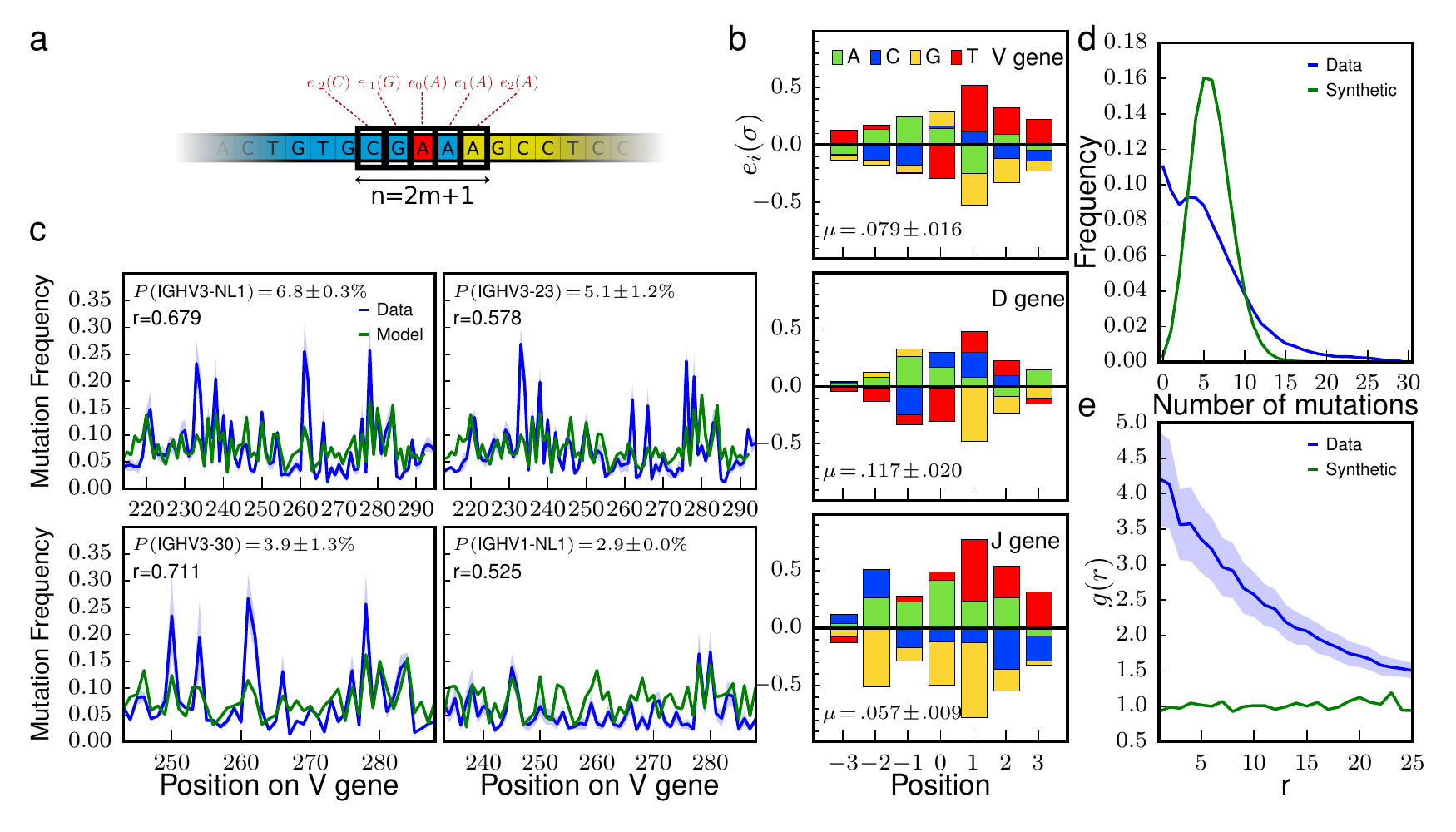}
\caption{
	\textbf{Hypermutation landscape.}
	(\textbf{a}) Position-Weight Matrix (PWM) model for predicting hypermutation hotspots in IGH. Each nucleotide $\sigma$ at position $i$ within $\pm m$ of the hypermutation site (in red) has an additive contribution $e_i(\sigma)$ to the hypermutation log odd (Eq.~\ref{mut}). The PWM is learned by Expectation-Maximization from the out-of-frame sequences of memory B cells.
	(\textbf{b}) Comparison between the observed mutation rate per nucleotide and its prediction by the PWM model, as a function of position along the V segment, for the four most frequent V genes. Pearson correlation coefficient $\rho$ and gene usage are given for each.
	(\textbf{c}) PWMs inferred from the V, D, and J genes.  
(\textbf{d}) Distribution of the number of mutations in each sequence. Data sequences have a broader distribution than predicted by the model (as computed from generating synthetic sequences and mutations with a data-inferred 7-mer PWM model).
(\textbf{e}) Spatial co-localization index $g(r)$, measuring the overrepresentation of pairs of hypermutations at genomic distance $r$ from each other. Synthetic sequences have $g(r)\approx 1$ by construction (green).
	}
\label{Fig5}
\end{figure*}
}

\makeatletter
\def\@seccntformat#1{%
  \expandafter\ifx\csname c@#1\endcsname\c@section\else
  \csname the#1\endcsname\quad
  \fi}
\makeatother

\begin{document}

\newcommand{\mytitle}{IGoR: a tool for high-throughput immune repertoire analysis}

\title{\mytitle}

\newcommand{\myauthors}{Quentin Marcou$^{1}$, Thierry Mora$^{*2}$ and Aleksandra M. Walczak$^{*1}$}

\author{\myauthors}

\affiliation{
	\normalsize{$^{1}$Laboratoire de Physique Th\'eorique, UMR8549,}\\
	\normalsize{CNRS and \'Ecole Normale Sup\'erieure, 24, rue Lhomond, 75005 Paris, France;}\\
	\normalsize{$^{2}$ Laboratoire de physique statistique, UMR8550,}\\
	\normalsize{CNRS, UPMC and \'Ecole normale sup\'erieure, 24,
          rue Lhomond, 75005 Paris, France.}\\
\normalsize{${}^*$ Equal contribution.}\\
	}

\date{\today}

\begin{abstract}
High throughput immune repertoire sequencing is promising to lead to new statistical diagnostic tools for medicine and biology. Successful implementations of these methods require a correct characterization, analysis and interpretation of these datasets. We present IGoR -- a new comprehensive tool that takes B or T-cell receptors sequence reads and quantitatively characterizes the statistics of receptor generation from both cDNA and gDNA. It probabilistically annotates sequences and its modular structure can investigate models of increasing biological complexity for different organisms. For B-cells IGoR returns the hypermutation statistics, which we use to reveal co-localization of hypermutations along the sequence. We demonstrate that IGoR outperforms existing tools in accuracy and estimate the sample sizes needed for reliable repertoire characterization. 

 \end{abstract}

\maketitle


The adaptive immune system recognizes pathogens by binding their antigens to specific surface receptors expressed on T and B cells. The recent advent of high throughput immune repertoire sequencing (RepSeq) \cite{Warren2013a,Six2013,Woodsworth2013b,Georgiou2014a} gives us direct insight into the diversity of B-cell and T-cell receptor (BCR and TCR) repertoires with great potential to change the way we diagnose, treat and prevent immune system related disorders.
A growing number of algorithms and software tools have been designed to address the new challenges of RepSeq, in particular sequence analysis, germline assignment and clone construction \cite{Brochet2008,Thomas2013,Gupta2015,bolotin2015mixcr,Duez2016,ralph_consistency_2016}.
However, each receptor sequence can be generated in a large number of ways, or ``scenarios,'' through recombination of genomic segments, insertions and deletions and hypermutations. Standard assignments introduce systematic errors when describing this inherently stochastic process. Quantitatively characterizing the diversity and the biases of these mechanisms remains a challenge for understanding adaptive immunity and applying RepSeq for diagnostics.

We present a flexible computational method and software tool, IGoR (Inference and Generation of Repertoires), that processes raw immune sequence reads from any source (cDNA or gDNA) and learns unbiased statistics of V(D)J recombination and somatic hypermutations. Using these statistics,  for each sequence IGoR outputs a whole list of potential recombination and hypermutation scenarios, with their corresponding likelihoods. 
IGoR's performance at identifying the correct scenario is 2.5 times better than current state-of-the-art methods. IGoR used as a sequence generator produces an arbitrary number of randomly rearranged sequences with the same statistics as in the dataset. 
Applied to BCRs, IGoR learns a context-dependent hypermutation model to identify hotspots, which allows for a comprehensive analysis of the mutational landscape of BCRs.

\section{Results}
\subsection*{Probabilistic assignment of recombination scenarios}
V(D)J recombination selects two or three segments (V and J for TCR $\alpha$ and BCR lights chains; V, D, and J for TCR $\beta$ and BCR heavy chains) from a library of germline genes, and assembles them while deleting base pairs and inserting other non-templated ones at the junctions (Fig.~\ref{Fig1}a).
B cell receptors can further diversify through somatic hypermutations during affinity maturation. The recombination process is degenerate, as the same sequence can be generated in many different ways \cite{Venturi:2006hk}. IGoR starts by listing the possible recombination and hypermutation scenarios leading to an observed sequence in the dataset. 
It then assigns probability weights reflecting the likelihood of these scenarios.
As the example in Fig.~\ref{Fig1}a shows, explored scenarios can be very different yet have comparable contributions to the sequence likelihood. Since exploring all possible scenarios would be computationally too costly, IGoR restricts its exploration to the reasonably likely ones. Scenario exploration takes from 1 ms up to less than a second per sequence on a single CPU core, depending on the chain (see full distributions of runtimes in Fig.~\ref{SIfig_runtime}). Different recombination architectures and dependencies can be configured within IGoR by specifying dependencies between elementary events (gene choices, deletions, insertions, hypermutations) through an acyclic directed graph, or Bayesian network, as illustrated in Fig.~\ref{Fig1}b for the case of TCR $\beta$ chains (see Online Methods for the other used structures). 

\figone

IGoR functions according to three modes: learning, analysis, and generation (Fig.~\ref{Fig1}c). In the learning mode, IGoR infers the recombination statistics of large datasets of sequences using a Sparse Expectation-Maximization algorithm (see Online Methods). In the analysis mode, IGoR assigns recombination events to sequences in a probabilistic way, by outputing the most likely scenarios ranked by their probabilities, as well as the overall generation probability of the sequence. In the generation mode, IGoR outputs random sequences with specified statistics, e.g. learned from real datasets.

\subsection*{Inference of V(D)J recombination}

\figtwo

We used IGoR's learning mode to infer the accurate statistics of V(D)J recombination from four datasets comprised of unique sequences of non-productive rearrangements of three different chains, sequenced either at the levels of mRNA (TCR$\alpha$ chain or TRA, and TCR$\beta$ chain or TRB \cite{pogorelyy2016persisting}) or DNA (TRB \cite{robins2009comprehensive}, BCR heavy chain or IGH from naive cells \cite{larimore_shaping_2012}), generalizing earlier methods \cite{Murugan2012,elhanati_inferring_2015,Elhanati2016}. Restricting to nonproductive unique sequences allowed us to avoid biases introduced by functional selection.
The Expectation-Maximization algorithm converged within a few iterations (see Fig.~\ref{SIFig_conv} for convergence of parameters, and Fig.~\ref{SIconvergence_fig} for the case of IGH).

The same TRB insertion and deletion distributions 
were inferred regardless of the individual, laboratory of origin, or sequencing protocol, and of whether DNA \cite{robins2009comprehensive} (light blue distributions in Fig.~\ref{Fig2}) or mRNA \cite{pogorelyy2016persisting} (dark blue) was used.
{By contrast, V and J gene usage varied moderately but significantly across individuals, and even more across sequencing technologies, suggesting possible primer-dependent biases (Fig.~\ref{SIFig_TCRmRNADNA}, see also Fig.~\ref{SIFig_BCR_DJ} for IGH D-J gene usage)}.
Insertions at the TRA V-J junction, and at the TRB V-D and D-J junctions have similar distributions (Fig.\ref{Fig2}a), as previously reported \cite{Elhanati2016}. IGH have significantly more insertions at the junctions than TCRs, consistent with previous observations \cite{elhanati_inferring_2015}.

\figthree

We then validated the learning algorithm on synthetic datasets. Sequences were generated in batches of $10^3$ to $10^5$ by IGoR with a variable error rate, using statistics inferred from {60bp DNA} TRB data. IGoR's learning algorithm was then run on these raw sequences, and the resulting statistics compared to the known ground truth.
We found that the inference was highly accurate for datasets of $10^5$ sequences and an error rate set to its typical experimental value, $10^{-3}$ (Fig.~\ref{Fig3}a and b), and was not affected by overfitting. However, not all high-throughput sequencing datasets reach this depth, especially when restricted to unique non-productive sequences. In addition, hypermutation rates in BCRs, which IGoR treats in the same way as errors, can reach 1-10\%. To assess how these limitations affect accuracy, we calculated the Kullback-Leibler divergence (a non-parametric measure of difference between probability distributions, see Online Methods) between the true distributions and the inferred ones, for varying sizes of datasets and error rates. For an error rate of $10^{-3}$, $\sim 5000$ unique out-of-frame sequences (which can be obtained from less than 2ml of blood with current mRNA sequencing technologies \cite{pogorelyy2016persisting}) were sufficient to learn an accurate model of TRB (Fig.~\ref{Fig3}c), with the
majority of the estimation error due to deletion profiles (which account for the majority of parameters).
Increasing the error rate has little effect up to rates of $10^{-2}$, but significantly degrades accuracy for typical hypermutation rates, $10^{-1}$ (Fig.~\ref{Fig3}d),
with the gene usage distribution affected the most (Fig.~\ref{SISampling_DKL_breakup}). This suggests that the recombination statistics of BCRs should be inferred using sequences from naive, non hypermutated cells (as we did in Fig.~\ref{Fig2}).

\subsection*{Analysis of scenario degeneracy}

\figfour

By considering all possible recombination scenarios for each sequence, our approach departs significantly from most existing methods, whose goal is to find the most likely one. To assess how often the most plausible scenario is the correct one, we analyzed synthetic sequences for which the generation scenario is known.
For each generated sequence, we used IGoR's analysis mode to enumerate the set of scenarios that were consistent with the nucleotide sequence, and ranked them according to their likelihood. 
Fig.~\ref{Fig4}a shows the distribution of the rank of the true recombination scenario for TRB and IGH synthetic data. The maximum-likelihood scenario is not the correct one in $72\%$ of IGH sequences and $85\%$ of 60bp TRB sequences. The distributions have long tails, meaning that a substantial fraction of sequences have a very large recombination degeneracy.

We then estimated how many scenarios, ranked from most likely to least likely, were needed to explain a given fraction $f$ of the total sequence likelihood. 
The distributions of this number across {100,000} generated sequences are shown in Fig.~\ref{Fig4}b for various values of $f$ (see Fig.~\ref{SIFig_probasTCR} for the equivalent plot for TRB data). To enumerate the correct scenario with $f=95\%$ confidence requires to include at least 30 to 50 scenarios. 
This analysis indicates that many scenarios need to be considered to correctly characterize the generation process.

IGoR outputs the probability of generation of the processed sequences, by summing the probabilities of all their possible scenarios, which deterministic assignment methods cannot do. It was shown that this generation probability was predictive of sharing properties between healthy individuals \cite{Murugan2012,pogorelyy2016persisting}. This functionality could be used as a useful indicator of convergent recombination in studies attempting to identify antigen-specific or auto-immune related sequences from large clinical datasets.

\subsection*{Comparison to other methods}

We compared our method to two representative state-of-the-art algorithms: MiXCR \cite{bolotin2015mixcr}, an efficient assignment tool that finds the best matching germline genes, and Partis \cite{ralph_consistency_2016}, a BCR-specific tool that uses maximum likelihood to find the most plausible scenario.
130 base-pair IGH sequences were synthetized in silico from a data-inferred model using IGoR's generation mode. We then assigned recombination scenarios using MiXCR, Partis and IGoR, and compared them to the true scenarios with which sequences were generated. In IGoR's and Partis' case, the model parameters were learned from the generated dataset 
to mimick the analysis of real data.
Fig.~\ref{Fig4}c shows the performance of the three methods in assigning the correct scenario of recombination.
IGoR performs about $2.5$ times better than MiXCR and Partis in predicting the complete recombination scenario, as well as each of its individual components. Note that Partis does not include palindromic insertions, which both IGoR and MiXCR treat by appending a short palindromic sequence at the end of each germline segment; restricting the analysis to sequences generated without palindromic insertions makes Partis' performance comparable to that of MiXCR (Fig.~\ref{SIFig_no_Pnuc_assign.pdf}).

Next, we compared the recombination statistics learned by the three methods to the true statistics used to generate the data. For MiXCR and Partis, we built the distribution of recombination events assigned to each sequence, while for IGoR these distributions were inferred using Expectation-Maximization, as explained before. All three methods yield similar statistics for V and J gene usage and deletion profiles (see Fig.~\ref{SIFig_compsoft}). However, the dependency between D an J usage in TRB is correctly captured by IGoR but not by the other methods (Fig.~\ref{Fig4}d).
TRB D and J genes are organised in two clusters, one containing D1 followed by genes of the J1 family, the other containing D2 followed by genes of the J2 family. Because of this organisation, D2 cannot be recombined with genes from the J1 family  \cite{murphy2016janeway}. MiXCR assigns 20\% of impossible D2-J1 recombination events to sequences (note that Partis does not process TCRs). By constrast, IGoR correcly learns the rule by assigning zero frequency to these impossible D-J pairs. The same results are obtained directly on real data (see Fig.~\ref{SIFig_TCR_DJ}).
Finally, IGoR accurately reconstructs the distribution of insertions, while the other methods systematically overestimate the probability of zero insertions (Fig.~\ref{SIFig_compsoft}a and b). 

\subsection*{Somatic hypermutations}

\figfive

To study patterns of SHMs in BCR expressed by memory B cells,
we included into
IGoR the possibility to infer a sequence-dependent hypermutation rate. The probability of error or mutation at a given position on the nucleotide sequence is assumed to depend on its immediate n-mer context (see Fig.~\ref{Fig5}a),  through the logistic transformation of an additive score computed using a position weigth matrix (PWM), similar to binding energy motifs used to describe DNA binding sites \cite{berg1987selection}.
We ran IGoR on memory out-of-frame IGH sequences from Ref. \cite{larimore_shaping_2012} to learn 7-mer PWMs, as well as overall mutation rates (the geometric mean of the mutation rate over all possible 7-mers), while fixing the recombination statistics to those previously learned from naive sequences, using Expectation Maximization (see Online Methods). IGoR's probabilistic framework handles the degeneracy of sequence origin caused by convergent combinations of gene choices and hypermutations. The learning procedure differs crucially from Ref.~\cite{elhanati_inferring_2015}, where the hypermutation rate was uniform.
Three distinct PWMs were learned for V, D, and J templated regions (Fig.~\ref{Fig5}b). To validate our PWM and mutation rate learning algorithm, we generated synthetic data with hypermutations according to the model learned from the real dataset, and re-learned its parameters using IGoR, finding excellent agreement (Fig.~\ref{SIFig_hyperm_synth}).

The PWM prediction for the position-dependent probability of hypermutations correlated well with that actually observed in the sequences ($r= 0.7$ for V genes, see Fig.~\ref{Fig5}c and Fig.~\ref{SIFig_mut_freq_pred}). PWMs were very reproducible across the two tested individuals ($r=0.98$, Fig.~\ref{SIFig_hyperm_real}), indicating that the inference procedure is robust to the individual history of infections, and pointing to the universal nature of the SHM mechanism. By constrast, the inferred overall mutation rate differred by a two-fold factor between the two individuals, probably owing to
differences in age, past infections, or lifestyle (Fig.~\ref{SIFig_hyperm_real}). 
The motifs we found recapitulate previously reported hotspot motifs (positive values of the PWM) for every gene, including WR\underline{C}Y (or WR\underline{C}H \cite{rogozin2004cutting}) and W\underline{A} \cite{betz1993passenger, shapiro1999predicting} (W = {A or T}, Y = {C or T}, R = {G or A}; mutated position underlined), as well as
 cold-spot motifs albeit to a lesser extend (SY\underline{C}, where S = {C, G}) \cite{bransteitter2004biochemical}. 
In all three motifs, C and G are generally underrepresented, except for the mutated position in V and D genes where T is less mutated than others. We assessed the robustness of the model to n-mer length by learning PWMs of sizes ranging from 3 to 9 (Fig.~\ref{SIFig_all_logos}). The contributions of each relative position did not change substantially as a function of context length. Positions at least up to 4 nucleotides away from the mutation locus contribute to the motif. This could mean that the context dependence is broad, or alternatively that the motif model is indirectly capturing non-contextual effects.
Overall, the inferred PWMs give both a more detailed and more nuanced view of the rules that govern hotspot positions, and cannot be reduced to a few easily describable motifs.

Fig.~\ref{Fig5}b shows that the motifs differ substantially between V, D, and J genes. V-learned PWMs only moderately predict  J-gene hypermutation rates ($r=0.5$ versus $r=0.7$ for V-gene rates), and J-learned PWMs predict V-gene rates even worse ($r=0.24$, see Fig.~\ref{SIFig_mut_freq_pred}).
This disagreement indicates that predictions purely based on context-dependent motifs are insufficient to explain all of the variability in hypermutation probabilities, and that other mechanisms must be at play. The overall mutation rate was also different between germline genes, consistent with reports that the chromatin state affects hypermutation rates \cite{kenter_aid_2016,steele_somatic_2016,chandra_aid_2015}.

We then used the inferred PWM within IGoR to probabilistically call putative hypermutations in sequences. We first examined the distribution of the number of mutations in a sequence (Fig.~\ref{Fig5}d). The empirical distribution (red) is more skewed and has a longer tail than would be expected by assuming independent hypermutations in each sequence, as predicted by generating randomly hypermutated sequences with the inferred PWM (blue). This observation is consistent with the fact that different B cells have undergone a variable number of cycles of affinity maturation, resulting in differences in effective hypermutation rates. Second, we asked whether hypermutations co-localized within the same sequence, by calculating the enrichment of hypermutations at two positions as a function of their genomic distance (Fig.~\ref{Fig5}e). While this enrichment is 1 in synthetic sequences (since our model assumes that hypermutations are independent of each other), real data shows up to a 4-fold enrichment of hypermutations at nearby positions. This difference is consistent with the fact that AID can cause repairs of DNA over large regions \cite{Unniraman2007}. The typical distance at which the co-localization enrichment index decays gives an estimate for the length of these correlated regions of hypermutations, about 15 base pairs.

IGoR can in principle calculate the generation probability of any sequence. However, highly hypermutated sequences pose an additional challenge because the ancestral (unmutated) recombined sequence itself is sometimes not known with certainty. To overcome this issue, IGoR explores  for each sequence all possible recombination and hypermutation scenarios, and calculates the generation probability of each potential ancestral sequence. Using synthetic data, we checked that the generation probability of individual sequences is well predicted by this method ($r=0.97$, see Fig.~\ref{SIFig_Pgen} and Online Methods), and its distribution accurately reproduced (see Fig.~\ref{SIFig_Pgen_density}).

\section{Discussion}
By treating alignments of immune receptors to the germline probabilistically \cite{Murugan2012}, IGoR corrects for systematic biases in the estimate of V(D)J recombination statistics, and predicts recombination scenarios more accurately than previous methods. Its detailed analysis of recombination scenarios further reveals that, even with a perfect estimator, the scenario is incorrectly called in more than 70\% of sequences, suggesting caution when interpreting results from deterministic assignments.

Although we demonstrated its functions on human TRB and IGH, IGoR's flexible structure makes it applicable to any variable lymphocyte receptor (TCR or immunoglobulin) and species for which genomic data is available. Unlike Hidden Markov Model based methods (e.g. \cite{ralph_consistency_2016,Elhanati2016}), it can include a wide array of possible dependencies between the recombination events. It
can also be adapted to handle unusual or incomplete rearrangements (D-J rearrangments, DD2/DD3 rearrangements in TCR $\delta$ chains, hybrid TRA/TRD recombinations, etc.). IGoR can also help detect unusual rearrangement features by using its syntheticaly generated sequences as a control. For instance, rearrangements with tandem Ds have been reported \cite{larimore_shaping_2012}, but distinguishing them from random insertions can be challenging. To test this, we counted sequences with two $\geq$10-nt D segments in the data, and compared it with predictions from IGoR's synthetic sequences generated with a single D segment (see Online Methods). We found 5 times more double-D assignments in IGH data than in the control, validating the findings of  \cite{larimore_shaping_2012}. In contrast, the same analysis performed on TRB showed no 
significant presence of tandem Ds. Future versions of IGoR should include the possibility of including multiple D rearrangements. Note that IGoR does not find reversed Ds in IGH (Fig.~\ref{SIFig_revDs}).

IGoR infers recombination statistics from non-productive sequences only, but can do it with as few as 5000 sequences. Once a recombination model is learned for a given locus, IGoR can generate arbitrary numbers of synthetic sequences with the same statistics, which could be used as a control in disease-association studies, by helping to distinguish antigen-specific clonotypes from public sequences with high convergent recombination frequencies, and thus dispense with the need of a healthy control cohort.

Our analysis of hypermutations led us to infer distinct sequence motifs for mutation targets on the V, D and J segments of human IGH, in contrast with
previous approaches that assume a universal context model \cite{cui_model_2016}. Although our motifs were learned on short reads comprising only part of the V and J segments, analysis of synthetic sequences showed that motifs could be accurately learned from such short reads. Exploring their applicability to longer reads would be an interesting future direction. 
We further found that hypermutations tend to co-localize along the sequence. These results suggest that at least three effects determine hypermutation hotspots: the immediate DNA context of the hypermutation, as modeled by our sequence motifs, position-specific effects mediated by e.g.
chromatin configuration and histone modifications, and the co-occurence of nearby mutations. Future improvements of hypermutation target predictions will have to rely on a better understanding of the precise mechanisms of AID operation \cite{chandra_aid_2015}.

{\bf Software availability.} IGoR along with example datasets and pre-learned human TRA, TRB, and IGH models is available at \url{bitbucket.org/qmarcou/igor}.

{\bf Acknowledgements.} The work was supported by grant ERCStG n. 306312.

\section{Online Methods}
IGoR functions according to three modes: VDJ statistics learning, sequence analysis, and sequence generation. All modes rely on an explicit stochastic description of the recombination and hypermutation events. In the analysis and learning modes, each sequence is analysed by listing all possible recombination and hypermutation scenarios. The learning mode iterates the analysis mode by updating the model parameters according to an Expectation-Maximization algorithm.

\subsection*{Recombination model}
In all three modes, IGoR assumes that receptor sequences result from a recombination scenario comprising several stochastic elements --- choice of germline segments, deletions and insertions. These features are stochastic and share statistical dependencies with each other. For tractability, we assume that these dependencies can be represented by an acyclic graph, also called Bayesian network (see SI Text for details). This structure can be configured within IGoR's setup files. For the purpose of this study, we used the following dependency structures for the $\alpha$ chain of T cells (TRA):
\begin{equation}
	\begin{split}
		P^{\alpha}_{\rm recomb} = {}& P(V,J)P(\text{del}V|V)P(\text{del}J|J) \\
		& \times P(\text{insVJ}) \dprod^{\text{insVJ}}_i P_{\rm VJ}(n_i|n_{i-1}),
	\end{split}\label{alpha}
\end{equation}
and for the $\beta$ chain of T cell receptors (TRB) and heavy chain of B cell receptors (IGH):
\begin{equation}
\begin{split}
P^{\beta/H}_{\rm recomb} = {}& P(V,D,J) P(\text{del}V|V) \\
& \times P(\text{insVD}) P(\text{del}Dl,\text{del}Dr|D) \\
& \times P(\text{ins}DJ) P(\text{del}J|J) \\
& \times \dprod^{\text{insVD}}_i P_{\rm VD}(n_i|n_{i-1})\dprod^{\text{insDJ}}_i P_{\rm DJ}(m_i|m_{i-1}).
\end{split}\label{beta}
\end{equation} 
where $V,D,J$ denote the choice of germline genes, ${\rm del}V$, ${\rm del}J$ the number of deleted base pairs at the ends of the V and J segments, ${\rm del}Dl$, ${\rm del}Dr$ the number of deletions at the left and right ends of the D segments, insVJ, insVD, insDJ,  the numbers of insertions at each of the insertion sites (between V-J, or V-D and D-J), and $n_i,m_i$ the identities of the inserted base pairs. In the case of TRB, gene usage is further factorized as $P(V,D,J) =P(V)P(D,J)$

\subsection*{Context dependent hypermutation model}
When processing TCRs or naive BCRs, a constant error probability is assumed throughout the sequence. When processing memory BCRs, a context-dependent hypermutation model is assumed: at each position along the V, D, and J genes, a hypermutation occurs with probability $P_{\rm mut}$, with
\begin{equation}
	\frac{P_{\rm mut}}{1-P_{\rm mut}} = \mu \exp{\left( \dsum_{i=-m}^{m}{ e_i(\pi_i)}\right)},\label{mut}
\end{equation}
where $(\pi_{-m},\ldots,\pi_{m})$ is the $(2m+1)$-mer sequence context centered around the location of the mutation. The entries of the position weight matrix (PWM), $e_i(\pi)$, contribute additively to the motif, and $\mu$ is the overall hypermutation rate.

\subsection*{Alignment to germline and scenario listing}
In the analysis and learning modes, each sequence is first aligned to all possible germline genes retrieved from genomic databases (e.g. IMGT), using the Smith-Waterman algorithm \cite{Smith1981}. Only germline genes with alignment scores higher than an adjustable threshold are considered for further analysis (see SI Text for details). Possible scenarios are then listed by picking germline genes with an above-threshold alignment score, and by choosing a number of base pairs to further delete from the ends of their aligned parts. The base pairs located between the germline segments trimmed in this manner are called insertions, and alignment mismatches to the germline are called  errors or hypermutations. When the palindromic end of germline genes is not entirely deleted, the number of remaining palindromic base pairs are described as negative deletions. To allow for the possibility that the D segments be inserted in both directions in BCRs, we added the reverse complements of each D germline segment to the list of genomic templates.

\subsection*{Sequence analysis}
For each sequence in the dataset, the probability of possible scenarios are computed using the recombination probability of Eqs.~\ref{alpha} or \ref{beta}, multiplied by the probability of errors or hypermutations $P_{\rm err}$: $P_{\rm scenario}=P_{\rm recomb}\times P_{\rm err}$. Scenarios are then listed in order of decreasing probability. The sum of probabilities $P_{\rm recomb}\times P_{\rm err}$ of possible recombination and hypermutation events gives the probability of observation of that particular sequence read, $P_{\rm read}$. The probability that the pre-mutation sequence was generated by recombination, $P_{\rm gen}$, is defined as the sum of the probabilities $P_{\rm recomb}$ of scenarios leading to that sequence. Since the pre-mutation sequence is not known with certainty, we calculated an approximate generation probability $P_{\rm gen}$ as the geometric mean of $P_{\rm gen}$ of all possible unmutated sequences consistent with the read, weighted by their posterior probabilities, $P_{\rm gen}\times P_{\rm err}/P_{\rm read}$. Alternatively, we approximated $P_{\rm gen}$ by that of the most likely pre-mutation sequence.

To shorten computation times, only plausible scenarios are listed by IGoR. Scenarios are enumerated by exploring the nodes of a hierarchical decision tree, where each depth corresponds to the choice of a scenario feature. Branches of the tree are discarded if their total contribution to the sequence probability is upper-bounded to be below a certain threshold. Details of the procedure are given in the SI text.

\subsection*{Learning algorithm}
The learning algorithm infers the parameters of Eqs.~\ref{alpha} or \ref{beta}, as well as the error or hypermutation model parameters of Eq.~\ref{mut}, from a large datasets of unique sequences. It relies on the sequence analysis module, and follows an Expectation-Maximization procedure. Starting from an arbitrary (but reasonable) set of parameters, all sequences in the dataset are analysed as described above, producing a long list of scenarios associated with each sequence. We define the pseudo-log-likelihood as the weighted sum of the log-likelihoods of all scenarios of all sequences, where the weights are given by the conditional probabilities of scenarios given the sequence, $P_{\rm recomb}/P_{\rm read}$ (Expectation step). This pseudo-log-likelihood is then maximized with respect to the parameters of the log-likelihoods (Eqs.~\ref{alpha}-\ref{mut}), while keeping the weights fixed. The parameters are updated, and the procedure repeated, until convergence. Mathematical derivations of the update rules and details about Expectation-Maximization are given in the SI Text.

\subsection*{Validation of model inference}
To compare the model parameters $\theta_1$ inferred from synthetic data to the known model parameters $\theta_2$ from which these data were generated, we computed the Kullback-Leibler divergence between two probability distributions, $D(\theta_1\Vert\theta_2)=\sum_E P(E,\theta_1)\log [P(E,\theta_1)/P(E,\theta_2)]$, where the sum is over all scenarios $E$. $P(E,\theta)$ is computed using Eqs.~\ref{alpha} or \ref{beta}. This Kullback-Leibler divergence can be decomposed into additive contributions from each of the scenario features, as detailed in the SI text.

\subsection*{Datasets}
We applied the learning algorithm on the following publicly available datasets: TCR alpha and beta chains RNA datasets from \cite{pogorelyy2016persisting} are available on Sequence Read Archive (SRP078490); TCR beta chains 60 bp DNA datasets from \cite{Murugan2012} are available at \url{http://physics.princeton.edu/~ccallan/TCRPaper/data/}; naive and Memory BCR heavy chains DNA datasets from \cite{elhanati_inferring_2015,Larimore2012} are available at \url{http://physics.princeton.edu/~ccallan/BCRPaper/data/}.

\subsection*{Correlations between hypermutations}
To evaluate correlations between the occurence of hypermutations at close-by positions along the BCR sequence, we computed the radial disbribution function defined as:
$g(r)=(1/N_r) \sum_{V;(i,j)\in C_V(r)} f(i,j,V)/f(i,V)f(j,V)$, where $f(i,V)$ and $f(i,j,V)$ are the frequencies of hypermutations at position $i$, and at both positions $i$ and $j$, respectively, calculated from individual scenario statistics weighted by their posterior probabilities. $C_V(r)$ is the set of pairs of positions separated by $r$ that were observed a large enough number of times in gene $V$, and $N_r=\sum_V |C_V(r)|$.

\subsection*{Usage of tandem D segments}
In order to assess the occurrence of double D insertions during the VDJ recombination event of IGH or TRB, we computed the frequency with which one could align (with the Smith-Waterman algorithm) two non-overlapping Ds over least 10 nucleotides, between the best V and best J alignments. We then compared the frequency obtained for synthetically generated sequences, to that obtained for real sequencing data.

\bibliographystyle{pnas}

\beginsupplement
\onecolumngrid

\section{Appendices}

\subsection{Model definitions}

We start by giving the particular model structures used in this study. We then give a more general definition applicable to other general types of recombination products.

\subsubsection{Models for TRA, TRB and IGH}

We define a probabilistic model for each type of chain (e.g. $\alpha$, $\beta$, heavy, light) that describes the probability of each  recombination event $\vec{E}$ by the probabilities of the known elements of the recombination subprocess (gene choice, insertions, deletions at each of the junctions etc) for each chain, and assumes only the minimum correlations between the subprocesses needed to explain the correlations observed in the data. We model insertions as a Markov chain (the identity of an inserted nucleotide only depends on the previously inserted one) with a nonparametric  length distribution \cite{Murugan2012,elhanati_inferring_2015,pogorelyy2016persisting}. For each insertion site (X= VD and DJ for $\beta$ and heavy chains and X=VJ for $\alpha$ and light chains) we infer the probability of observing a non-templated sequence of a given length,  $P(\rm{insX})$, and the transition matrices $P_{\rm VJ}(n_i|n_{i-1})$,  $P_{\rm VD}(n_i|n_{i-1})$,  $P_{\rm DJ}(m_i|m_{i-1})$ giving the probability of inserting a given nucleotide as a function of the identity of previous one. For each gene we infer the probability of the number of deletions conditioned on the gene identity, e.g. $P(\text{del}V|V)$ for deletions from the V gene. We model templated palindromic insertions as negative deletions \cite{Murugan2012,elhanati_inferring_2015}. The D gene is very short and may get fully deleted. This introduces correlations between the deletions on both sides of the original D gene template. We account for these correlations by inferring the joint probability $P(\text{del}Dl,\text{del}Dr|D)$. We treat every allele as a different gene \cite{elhanati_inferring_2015} and infer the joint gene usage $P(V,D,J)$ for $\beta$ and heavy chains, and $P(V,J)$ for $\alpha$ and light chains, to be able to capture correlations between segment usage. 

For TCR  $\alpha$ chains or BCR light chains,  the probability of a recombination event $\vec{E}=(V,J,\text{del}V,\text{del}J,\text{insVJ})$ is:

\begin{equation}\label{alphal}
\begin{split}
P^{\alpha/{\rm L}}_{\rm recomb}(\vec{E}) = {}& P(V,J)P(\text{del}V|V)P(\text{del}J|J) \\
& \times P(\text{\rm{insVJ}}) \dprod^{ {{\rm{insVJ} } } }_i P_{\rm VJ}(n_i|n_{i-1})
\end{split}
\end{equation}

Similarly, the probability $P^{\beta/h}_{\rm recomb}(\vec{E})$ of a recombination event $\vec{E}=(V,D,J,\text{del}V,\text{del}Dl,\text{del}Dr,\text{del}J,\text{insVD},\text{ins}DJ)$ for a TCR$\beta$ or BCR heavy chain is: 
\begin{equation}\label{betah}
\begin{split}
P^{\beta/H}_{\rm recomb}(\vec{E}) = {}& P(V,D,J) P(\text{del}V|V) \\
& \times P(\text{insVD}) P(\text{del}Dl,\text{del}Dr|D) \\
& \times P(\text{ins}DJ) P(\text{del}J|J) \\
& \times \dprod^{{\rm{insVD}}}_i P_{\rm VD}(n_i|n_{i-1})\dprod^{{\rm{insDJ}}}_i P_{\rm DJ}(m_i|m_{i-1}).
\end{split}
\end{equation} 
In the case of TRB, gene usage is further factorized as $P(V,D,J) =P(V)P(D,J)$.

\subsubsection{General model formulation}\label{general_formulation}

IGoR is designed in a modular way so the user can define arbitrary model forms. The models are Bayesian networks encoded as directed acyclic graphs, whose vertices ${i=1,\ldots,K}$ label individual recombination subprocesses $E_i$ (V, D, J choices, deletions, etc. in the examples above). Dependence of the recombination process $j$ upon $i$ is encoded by a directed edge between $i$ and $j$, denoted $v_{ij}=1$ (while $v_{ij}=0$ means no direct dependence). The set of parents of $i$, i.e. processes on which $i$ depends directly, is denoted by $\mathcal{P}_i=\{j|v_{ji}=1\}$.

Using these definitions we can, generally and irrespectively of the assumed form of the underlying model of recombination, write the probability of a complete recombination scenario $\vec{E}=(E_{1},\dots,E_{K})$ as:
\begin{equation}
	P_\text{\rm recomb}(\vec{E}|\theta) = \prod_{i=1}^K P(E_{i}|\{E_j\}_{j\in \mathcal{P}_i},\theta),
\end{equation}
where $\theta$ denotes the underlying model parameters (i.e. probability distributions of gene choice, insertions at a given junction, and deletions from a given gene in the studied examples).

Each recombination scenario $\vec{E}$ leads to a unique sequence $\vec{\hat S}(\vec{E})=(\hat S_1,\ldots,\hat S_L)$, $\hat S_i(E)\in\{A,C,G,T\}$ (in the following we often write $\vec{S}$ for $\vec{\hat S}(\vec{E})$ for brevity). However, in order to produce a given sequence $\vec{S}$ several scenarios might be equivalent, and we can write the probability of generating a given sequence as:
\begin{equation}
	P_{\rm gen}(\vec{S}|\theta)=\sum_{\vec{E} | \vec{\hat S}(\vec{E})=\vec{S}}P_\text{\rm recomb}(\vec{E}|\theta).
\end{equation} 
The above description only holds to assess the generation probability of a pure product of recombination and does not account for sequencing errors or hypermutations. Note that, since longer reads allow for more reliable determination of V and J gene segments, $P_{\rm gen}$ depends in general on read length: shorter reads can be created in more ways than longer reads, leading to larger $P_{\rm gen}$.

\subsubsection{Errors and hypermutations}\label{perfect_seq}
Sequencing  is inherently noisy and introduces nucleotide substitutions. In addition, BCRs can accumulate hypermutations, which can be mathematically treated in the same way as errors. For the sake of clarity, we  distinguish between the sequencing {read} $\vec{R}$ and the original {sequence} $\vec{S}$ resulting from recombination, as defined above. For simplicity we ignore insertion and deletion errors, so that $\vec{R}$ and $\vec{S}$ are of the same length $L$. 

We define our error model as deviations from the initial recombination event (through sequencing errors or somatic hypermutations) such that $P_{\rm err}(\vec{R}|\vec{S},\theta)$ is the probability of observing the sequencing read $\vec{R}$ given the recombination product $\vec{S}$. Since the recombination scenario $\vec{E}$ completely determines $\vec{S}$, $P_{\rm err}(\vec{R}|\vec{S},\theta) = P_{\rm err}(\vec{R}|\vec{E},\theta)$, and we use these two notations interchangeably. The dependence on $\theta$ reflects the fact that $\theta$ also includes the parameters of the error or hypermutation model.

We write the joint probability of producing a given sequence $\vec{S}$ and observing a given read $\vec{R}$ as:
\begin{equation}
	P(\vec{R},\vec{S}|\theta) = P_{\rm gen}(\vec{S}|\theta) P_{\rm err}(\vec{R}|\vec{S},\theta).
\end{equation}
Summing over all possible recombination products, the likelihood of a sequencing read is:
\begin{equation}
\begin{split}
	P_{\rm read}(\vec{R}|\theta)  & = \dsum_{\vec{S}}P(\vec{R},\vec{S}|\theta) \\
	& =  \dsum_{E}P_\text{\rm recomb}(E|\theta)P_{\rm err}(\vec{R}|\vec{E},\theta),
\end{split}
\end{equation}
and the total likelihood of the model given a dataset of reads $(\vec{R}^1,\ldots,\vec{R}^N)$ is given by:
\begin{equation}
	\mathcal{L}_{\rm total}(\theta) = \prod_{a=1}^N P_{\rm read}(\vec{R}^a|\theta).
\end{equation}

\subsection{Expectation-maximization}

\subsubsection{General scheme}

The recombination machinery is degenerate, as several scenarios of recombination and hypermutations can lead to the same sequence, and the recombination scenario $\vec{E}$ from which the sequencing read $\vec{R}$ comes from is in general unknown. The Expectation-Maximization algorithm is a commonly used algorithm that maximizes the likelihood of models with hidden variables given the data. In this section we re-derive this algorithm for our class of models. 

The procedure is iterative. Starting from an initial set of parameters $\theta$, one wishes to update another set of parameters $\theta'$.
From Bayes formula, $P_{\rm read}(\vec{R}|\theta')={P(\vec{E},\vec{R}|\theta')}/ P(\vec{E}|\vec{R},\theta')$, we rewrite the log-likelihood of a read as:
\begin{equation}
\ln P_{\rm read}(\vec{R}|\theta')=\sum_{\vec{E}} P(\vec{E}|\vec{R},\theta) \left[\ln{P(\vec{E},\vec{R}|\theta')}-\ln P(\vec{E}|\vec{R},\theta')\right]=  q(\theta'|\theta,\vec{R}) + h(\theta'|\theta,\vec{R}),
\end{equation}
where we have used $\sum_\vec{E}{P(\vec{E}|\vec{R},\theta)}=1$, and where we have defined
\begin{align} h(\theta'|\theta,\vec{R})&=-\sum_\vec{E}{P(\vec{E}|\vec{R},\theta)}\ln{P(\vec{E}|\vec{R},\theta')},\\ q(\theta'|\theta,\vec{R})&=\sum_\vec{E}{P(\vec{E}|\vec{R},\theta)}\ln{P(\vec{E},\vec{R}|\theta')}.
\end{align}

The difference between the log-likelihood, $\ln\mathcal{L}_{\rm total}(\theta)=\sum_{a=1}^N \ln P_{\rm read}(\vec{R}|\theta)$, between the current set of parameters $\theta$ and the candidate new parameters $\theta'$ reads:
\begin{equation}
	\begin{split}
\ln\mathcal{L}_{\rm total}(\theta')-\ln\mathcal{L}_{\rm total}(\theta)&=\sum_{a=1}^N
		 q(\theta'|\theta,\vec{R}^a) - q(\theta|\theta,\vec{R}^a) 
		 + h(\theta'|\theta,\vec{R}^a) - h(\theta|\theta,\vec{R}^a).\\
&\geq \sum_{a=1}^N
		 q(\theta'|\theta,\vec{R}^a) - q(\theta|\theta,\vec{R}^a) \\
&\geq 		 Q(\theta'|\theta) - Q(\theta|\theta) 
	\end{split}
\end{equation}
where $Q(\theta'|\theta)=\sum_{a=1}^N q(\theta'|\theta,\vec{R}^a)$, and where we have used Gibbs inequality:
\begin{equation}
	h(\theta'|\theta,\vec{R}^a) - h(\theta|\theta,\vec{R}^a) =\sum_\vec{E}{P(\vec{E}|\vec{R},\theta)}\ln\frac{P(\vec{E}|\vec{R},\theta)}{P(\vec{E}|\vec{R},\theta')}\geq 0.
\end{equation}

This inequality ensures that maximizing the ``pseudo-log-likelihood''  $Q(\theta'|\theta)$ over $\theta'$ increases total likelihood by at least the same amount. The Expectation-Maximization scheme updates $\theta$ by doing such a maximization, and repeating the procedure iteratively. The algorithm converges to a maximum of the likelihood.

\subsubsection{Optimizing the recombination model}
The pseudo-log-likelihood can be broken up in two independent terms, 
$Q(\theta'|\theta) = Q_{\text{\rm recomb}}(\theta'|\theta) + Q_{\text{err}}(\theta'|\theta)$, respectively corresponding to the recombination model and the error or hypermutation model:
\begin{align}
	Q_{\text{\rm recomb}}(\theta'|\theta) &= \sum_{a=1}^N \sum_{\vec{E}} {P(\vec{E}|\vec{R}^a,\theta)} \ln {P_\text{\rm recomb}(\vec{E}|\theta')}.\\
Q_{\rm err}(\theta'|\theta)&=\sum_{a=1}^N \sum_{\vec{E}} {P(\vec{E}|\vec{R}^a,\theta)} \ln {P_\text{\rm err}(\vec{R}|\vec{E},\theta')}.
\end{align}

In order to maximize the pseudo-log-likelihood of the recombination model we need to maximize $Q_{\text{\rm recomb}}(\theta'|\theta)$ with respect to every model component contained in the parameter set $\theta'$, $P'(E_{i}|\{E_j\}_{j\in \mathcal{P}_i})$. We impose normalization using Lagrange multipliers, $\lambda_i$, and define:
\begin{equation}
	\hat{Q}_{\text{\rm recomb}}(\theta'|\theta) = Q_{\text{\rm recomb}}(\theta'|\theta) + \sum_{i}\lambda_i\left [1-\sum_{E_i} P'(E_i|\{E_j\}_{j\in \mathcal{P}_i})\right].
\end{equation}

Taking the functional derivative of $\hat{Q}_{\text{\rm recomb}}(\theta^*|\theta)$ with respect to the model parameter we get:
\begin{equation}
\frac{\partial \hat{Q}_{\text{\rm recomb}}(\theta'|\theta)}{\partial P'(E_i|\{E_j\}_{j\in \mathcal{P}_i})} = \sum_{a=1}^N \sum_{\vec{E}'} \delta_{E_i,E'_i}\frac {P(\vec{E}'|\vec{R}^a,\theta)}{P'(E_i|\{E_j\}_{j\in \mathcal{P}_i})} + \lambda_i.
\end{equation}

Setting this derivative to zero gives:
\begin{equation}\label{eq21}
	P'(E_i|\{E_j\}_{j\in \mathcal{P}_i}) = \frac{1}{N}\sum_{a=1}^N \sum_{\vec{E}'} \delta_{E_i,E'_i}P(\vec{E}'|\vec{R}^a,\theta),
\end{equation}
where the Lagrange parameter $\lambda_i=N$ ensures normalization.
In other words the modified log-likelihood is maximized by using an update rule that equates the probability of a realization of a recombination event to its posterior frequency.

\subsubsection{Optimizing the independent single nucleotide error model}
The independent single nucleotide error model is the simplest instance of an error model, where each nucleotide of the read has a probability $r$ to be mis-sequenced as one of the three other nucleotides with equal probability. For this model we have
\begin{equation}
	\begin{split}
		P_{\rm err}(\vec{R}|\vec{S},\theta) = \left(\frac{r}{3}\right)^{N_{\rm err}} (1-r)^{L)-N_{\rm err}(\vec{R},\vec{S})}.
	\end{split}
\end{equation}
where $N_{\rm err}(\vec{R},\vec{S})$ the number of mismatches between $\vec{R}$ and $\vec{S}$, and $L$ the number of error-prone base pairs.
We compute the derivative of the modified log-likelihood of the error model with respect to $R^*$ as:
\begin{equation}
	\frac{d Q_{\text{err}}(\theta'|\theta)}{d r'} = 
	\sum_{a=1}^N\sum_{\vec{E}}P(\vec{E}|\vec{R}^a,\theta)\left( \frac{N_{\rm err}(\vec{R}^a,\vec{\hat S}(\vec{E}))}{r'} - \frac{L(\vec{R}^a,\vec{E})-N_{\rm err}(\vec{R}^a,\vec{\hat S}(\vec{E}))}{1-r'} \right).
\end{equation}

Setting this derivative to zero yields:
\begin{equation}
	R' =\frac{\sum_{a=1}^N\sum_{\vec{E}}P(\vec{E}|\vec{R}^a,\theta)N_{\rm err}(\vec{R}^a,\vec{\hat S}(\vec{E}))}{\sum_{a=1}^N \sum_{\vec{E}}P(\vec{E}|\vec{R}^a,\theta) L(\vec{R}^a,\vec{E})},
\end{equation}
where $L(\vec{R}^a,\vec{E})$ is the number of potentially erroneous nucleotides in read $a$.
For simplicity we ignore errors and hypermutations in the insertion part of the sequence, as they are almost indistinguishable from unmutated random insertions, and accounting for them would imply summing over an exponentially large number of scenarios.
As a result, $L$ in the above formula is not the read length, but rather the number of genomic nucleotides in each scenario, which depends on the scenario $\vec{E}$ as well as on the sequence read. 

\subsubsection{Optimizing the hypermutation model}
The hypermutation model assumes the following form for the probability of hypermutations:
\begin{equation}
P_{\rm err}(\vec{R}|\vec{S})=\prod_{x,S_x\neq R_x}\frac{P_{\rm mut}(S_{x-m},\ldots,S_{x+m})}{3}\prod_{x,S_x=R_x} \left[1-P_{\rm mut}(S_{x-m},\ldots,S_{x+m})\right],
\end{equation}
with
\begin{equation}
	\frac{P_{\rm mut}(\vec{\pi})}{1-P_{\rm mut}(\vec{\pi})} = \mu \exp{\left( \sum_{i=-m}^{m}{ e_i(\pi_i)}\right)},
\end{equation} 
where $(\pi_{-m},\ldots,\pi_{m})=(S_{x-m},\ldots,S_{x+m})$ is the sequence context of the original recombination product around a hypermutation at position $x$. The parameters $e_i(N)$, the position-weight matrix, and $\mu$, the overall mutation rate, are part of the parameter set $\theta$.
In order to lift the degeneracy of the model we impose that $\sum_{N={A,C,G,T}} e_i(N) = 0$ at every position $i$.

The pseudo-log-likelihood of the hypermutation model reads:
\begin{equation}
Q_{\text{err}}(\theta'|\theta) = \sum_{a=1}^M\sum_{\vec{E}}P(\vec{E}|\vec{R}^a,\theta) 
\sum_{x=1}^{L} \left[ \delta_{S_x,R_x} \ln \frac{1}{1+r'(\vec{S},x)}  + (1-\delta_{S_x,R_x})\ln \frac{r'(S,x)/3}{(1+r'(\vec{S},x))} \right], 
\end{equation}
where $r'(S,x) =r'(S_{x-m},\ldots,S_{x+m})= \mu' \exp{ \left (\sum_{i=-m}^{m}{e'_i(S_{x+i})} \right )}$. It can be rewritten as:
\begin{equation}
		Q_{\text{err}}(\theta'|\theta) = \sum_{\vec\pi} \left[\left(\ln(\mu'/3) + \dsum_{i=0}^N{e_i'(\pi_i)} \right)N_{\text{mut}}(\vec\pi)
			 - \ln\left(1+\mu' \exp{\left( \dsum_{i=1}^N{ e'(\pi_i)}\right)}\right)N_{\text{bg}}(\vec\pi)\right],
\end{equation}
where
\begin{align}
N_{\text{bg}}(\vec\pi) & =  \sum_{a=1}^M \sum_{\vec{E}}P(\vec{E}|\vec{R}^a,\theta) 
\sum_{x=1}^{L}  \prod_{i=-m}^m \delta_{S_{x+i},\pi_i}\\
N_{\text{mut}}(\vec\pi) & = \sum_{a=1}^M\sum_{\vec{E}}  P(\vec{E}|\vec{R}^a,\theta) 
\sum_{x=1}^{L}  (1-\delta_{S_x,R_x}) \prod_{i=-m}^m \delta_{S_{x+i},\pi_i}.
\end{align}

During the Expectation step, we compute these two quantities for each (2m+1)-mer and then maximize $Q_{\rm err}$ at each step of the Expectation-Maximization scheme using Newton's method with a backtracking line search. To impose $\sum_{\sigma} e_i(\sigma) = 0$ we remove one parameter per position $i$ by setting for one nucleotide, $e_i(N) = -\sum_{\sigma \neq N} e_i(\sigma) $.

We can then compute the entries of the gradient vector $\vec{J}$ (of size $3(2m+1)+1$):
\begin{align}
	\frac{\partial Q_{\text{err}}(\theta'|\theta)}{\partial \mu'} & = \dsum_{\vec{\pi}} \left( \frac{N_{\rm mut}(\vec{\pi})}{\mu'} - N_{\rm bg}(\vec{\pi}) \frac{r'(\vec{\pi})}{\mu'(1+r'(\vec\pi))} \right),
	\\
	\frac{\partial Q_{\text{err}}(\theta'|\theta)}{\partial e_i'(\sigma)}& = \dsum_{\vec{\pi}}(\delta_{\vec{\pi}_i,\sigma}-\delta_{\vec{\pi}_i,N})\left[ N_{\rm mut}(\vec{\pi}) - N_{\rm bg}(\vec{\pi})\frac{r'(\vec\pi)}{1+r'(\vec\pi)} \right] ,
\end{align}
along with the Hessian matrix $\vec{H}$ entries:
\begin{align}
	\frac{\partial^2 Q_{\text{err}}(\theta'|\theta)}{\partial \mu'^{2}} & = \dsum_{\vec{\pi}} \left( N_{\rm bg}(\vec{\pi})\frac{r'(\vec{\pi})^2}{\mu'^{2}(1+r'(\vec{\pi}))^2} - \frac{N_{\rm mut}(\vec{\pi})}{\mu'^{2}} \right),
	\\
	\frac{\partial^2 Q_{\text{err}}(\theta'|\theta)}{\partial \mu' \partial e'_i(\sigma)} & = \dsum_{\vec{\pi}}(\delta_{\vec{\pi}_i,N}-\delta_{\vec{\pi}_i,\sigma})N_{\rm bg}(\vec{\pi})\frac{r'(\vec{\pi})}{\mu'(1+r'(\vec{\pi}))^2},
	\\
		\frac{\partial^2 Q_{\text{err}}(\theta'|\theta)}{\partial e'_i(\sigma)\partial e'_j(\sigma')}  &= \dsum_{\vec{\pi}}(\delta_{\vec{\pi}_i,N}-\delta_{\vec{\pi}_i,\sigma})(\delta_{\vec{\pi}_j,N}-\delta_{\vec{\pi}_j,\sigma'})
		 N_{\rm bg}(\vec{\pi})\frac{r'(\vec{\pi})}{(1+r'(\vec{\pi}))^2}.
\end{align}

For each step of Newton's method we find the step direction by solving $\vec{H}\Delta\theta' = -\vec{J} $ and we gradually refine the step size based on the Armijo-Goldstein condition. These operations are iteratively repeated until the pseudo-log-likelihood of the error model for a given Maximization step of the EM framework is maximized. 

\subsection{Model entropy and $D_{\textbf{KL}}$}
Shannon's entropy \cite{shannon1948mathematical,cover2012elements},
\begin{equation}
S(\theta)=\sum_x p(x|\theta)\ln p(x|\theta),
\end{equation}
is a measure of the uncertainty about the outcome of a stochastic process described by a variable  $x$, governed by the distribution $p(x|\theta)$ and parametrized by $\theta$. As in \cite{elhanati_inferring_2015,Murugan2012,Elhanati2016} we compute this quantity based on our probabilistic framework and use it as an estimate for the diversity generated by the V(D)J recombination process. 
In the main text we also introduced the relative entropy or Kullback-Leibler divergence,
\begin{equation}
D(\theta_1||\theta_2) = \sum_x p(x|\theta_1)\ln\frac{p(x|\theta_1)}{p(x|\theta_2)},
\end{equation}
as a measure of dissimilarity between two probability distributions parametrized by $\theta_1$ and $\theta_2$ respectively, and used it to quantify the error made by our probabilistic framework upon inferring the V(D)J recombination parameters.

Since both the entropy and the Kullback Leibler divergence between two recombination models can be computed once one knows how to compute the cross entropy $H(\theta_1,\theta_2) = \sum_x p(x|\theta_1)\ln p(x|\theta_2)$ between the distributions for the two sets of parameters $\theta_1$ and $\theta_2$, we focus here on the computation of $H(\theta_1,\theta_2)$.
 
\subsubsection{General form}
For the considered class of models, the cross-entropy can be divided into subparts for each model component, 
\begin{equation}
H(\theta_1,\theta_2) = \sum_{i=1}^KH_i(\theta_1,\theta_2),
\end{equation}
with
\begin{equation}
	H_i(\theta_1,\theta_2) = \sum_{\vec{E}}P(\vec{E}|\theta_1)\ln P(E_i|\{E_j\}_{j\in \mathcal{P}_i},\theta_2).
\end{equation}
To calculate this sum, one does not need to sum over all possible scenarios $\vec{E}$, but only on combinations of processes that affect $E_i$ directly or indirectly. Let us call $A_i\subset \{1,\ldots,K\}$ the set of indices affecting process $i$. These are defined as the ``ancestors'' of $i$ in the acyclic graph, i.e. indices $j$ such that there exists a lineage from $j$ to $i$, $(i_1=i,i_2,\ldots,i_k=j)$ with $i_{\ell+1}\in \mathcal{P}_{i_{\ell}}$ (note that $A_i$ includes $i$ itself as a 0th order ancestor). Then the previous sum can be reduced to a sum over the processes in $A$ only:
\begin{equation}
	H_i(\theta_1,\theta_2) = \sum_{\vec{E}_{A_i}}\left[\prod_{j\in A_i}P(E_j|\{E_{j'}\}_{j'\in \mathcal{P}_j},\theta_1)\right]\ln P(E_i|\{E_j\}_{j\in \mathcal{P}_i},\theta_2).
\end{equation}
where $\vec{E}_{A_i}$ denotes the subvector of elements of $\vec{E}$ with indices in $A$.
Estimating the cross entropy for an event $E_i$ requires exponential time in the number of ancestors of that node. Fortunately, in the recombination models considered in this paper the set of ancestors are small and obtaining the cross entropy is easy for every event. The special case of insertions is discussed below.
Note that this cross-entropy only takes into account the recombination model, and not the error model.  

\subsubsection{Inserted nucleotides}
For a given insertion length insVJ (or insVD, or insDJ), the cross-entropy between two models of insertions is given by
\begin{align}
	h({\rm insVJ},\theta_1,\theta_2) &= \sum_{\vec{n}} P(\vec{n},\theta_1)\ln P(\vec{n},\theta_2)\\
&=\sum_{n_1}P_{s}(n_1|\theta_1)\ln P_s(n_1|\theta_2)\\&+({\rm insVJ}-1)\sum_{n_1,n_2}P_{s}(n_1|\theta_1) P(n_2|n_1,\theta_1)\ln P(n_2|n_1,\theta_2)
\end{align}
where $\vec{n}=(n_1,\ldots,n_{\rm insVJ})$ is the inserted sequence, and $P_s(n_1,\theta)$ is the stationary distribution of the Markov chain of insertions, solution of the equation $\sum_{n_0}P(n_1|n_0,\theta)P_s(n_0,\theta)=P_s(n_1,\theta)$.
The average cross-entropy over possible lengths is then given by:
\begin{equation}
	H_{\rm VJ~insertions}(\theta_1,\theta_2) =  \sum_{\vec{E}_{B}}\left[\prod_{j\in B}P(E_j|\{E_{j'}\}_{j'\in \mathcal{P}_i},\theta_1)\right]h({\rm insVJ},\theta_1,\theta_2),
\end{equation}
 where $B\subset \{1,\ldots,K\}$ is the subset of processes affecting either  insVJ or $\vec{n}$, exluding insVJ itself.

\subsection{Probability of generation}
Although the probability of generation of a sequence without errors or hypermutations is well defined, computing the probability of generation of a mutated sequence, before mutations occurred, is strictly speaking not possible because that sequence is not know with certainty. However, we can compute a good approximation for it, and we can also calculate its distribution across sequences.

To approximate $P_{\rm gen}(\vec{S})$ from a noisy or hypermutated sequence $\vec{R}$, we take its geometric average weighted by the probability of the recombination product $\vec{S}$:
\beq
\ln P^*_{\rm gen}(\vec{R})\approx \sum_{\vec{E}} P(\vec{E}|\vec{R},\theta)\ln P_{\rm gen}(\vec{\hat S}(\vec{E}),\theta),
\eeq
with $P(\vec{E}|\vec{R},\theta) =P_{\rm recomb}(\vec{E},\theta)P_{\rm err}(\vec{R}|\vec{\hat S}(\vec{E}),\theta)/P_{\rm read}(\vec{R},\theta)$. Alternatively, one can take the generation probability of the most likely recombination product:
\beq
P^*_{\rm gen}(\vec{R})\approx P_{\rm gen}(\vec{S}^*,\theta),
\eeq
where $\vec{S}^*=\argmax_{\vec{S}} P(\vec{S}|\vec{R},\theta)$.

The distribution $\rho(x)$ of the log-probabilities of generation, $x=\log P_{\rm gen}$, can be computed from data using:
\beq
\rho(x)=\frac{1}{N}\sum_{a=1}^N \sum_{\vec{E}} P(\vec{E}|\vec{R},\theta) \delta\left[x-\ln P_{\rm gen}(\vec{\hat S}(\vec{E}),\theta)\right].
\eeq
Note that unlike estimates for single sequences, this expression should become exact in the limit of $N\to\infty$.

\subsection{Data and software}
\subsubsection{Genomic templates}
We used custom genomic templates derived from the IMGT database \cite{lefranc2009imgt}.
TCR alpha V and J genomic templates were taken from the IMGT human database.
For TCR beta V, D and J genes we used curated genomic templates from \cite{Murugan2012}.
BCR heavy chain V, D and J genes were taken from the customized genomic templates used in \cite{elhanati_inferring_2015}.
For software comparison we used default genomic templates provided with Partis and MiXCR.

\subsubsection{Alignments}
Initial alignments to germline genes were performed using the Smith-Waterman algorithm \cite{Smith1981}, with scores of 5 for matching base pairs, -14 for mismatches, and a 50 gap penalty. Alignments with a score below the following gene dependent threshold were discarded: 50 for TRBV, 0 for TRBD, 10 for TRBJ, 20 for TRAV, 10 for TRAJ, 50 for IGHV, 40 for IGHD, 10 for IGHJ. We also discarded alignments whose score fell below the maximum alignment score (found for this read and segment type), minus the following variable range: 55 for TRBV, 35 for TRBD, 10 for TRBJ, 55 for IGHV, 20 for IGHJ.

The alignment offset (the index of the nucleotide on the read to which the first letter of the undeleted genomic template is aligned) was constrained depending on known primer locations on the J gene.

\subsubsection{Pruning the tree of scenarios}
Since enumerating all possible scenarios for each sequence is not tractable, we used a heuristic method for reducing their numbers. Exploring all possible scenarios is equivalent to exploring all the terminal leafs of a tree. Our heuristic is to prune all branches that do not contribute substantially to the likelihood of the read. To do this we implement a Sparse Expectation Maximization algorithm as motivated in \cite{neal1998view}. {{Due to the acyclicity of the directed graph underlining the Bayesian network, there exists a topological sorting of the events constituting a partially ordered set (we will assume in the following that the indices of the different events $E_i$ respect this ordering)}.  IGoR processes event realizations  according to this order corresponding to different layers of depth in the tree. To discard irrelevant branches (containing negligible scenarios) IGoR computes at each depth $k$ (with $0\leq k < K$) an upper bound on the probability of the currently explored scenario:

\begin{equation}
	\label{eq_tree_pruning}
	\frac{\dprod_{0 \leq i \leq k}P( E_i,\vec{R}|\{E_j\}_{j\in\mathcal{P}_i},\theta)\dprod_{k<i<K}{\underset{e_i}{\rm{max}}~P( E_i,\vec{R}|\theta)}}{\underset{\vec{E} \in \mathcal{E}}{\rm max}~P(\vec{E},\vec{R}|\theta)} >\varepsilon,
\end{equation}
where $\mathcal{E}$ is the set of already fully explored scenarios, and $0 \leq \varepsilon \leq 1$ is a tunable parameter setting the precision of the sparsity approximation. While $\varepsilon=0$ will explore every possible scenario and perform an exact Expectation step, $\varepsilon=1$ will explore only scenarios more likely that any scenario already explored.

Although Eq.~\ref{eq_tree_pruning} captures the essence behind our tree pruning approach, in practice IGoR uses more information than a simple upper probability bound. By picking two gene choice realizations, imposing the identity and position of these specific V and J genes, we explicitly impose the total nucleotide length of event realizations between those V and J genes (number of insertions, deletions, D gene length, ...). When computing the probability upper-bounds IGoR computes the upper probability bound for a given junction length between two event realizations, and uses this refined bound to efficiently prune the tree of scenarios. 

\subsubsection{Generating synthetic sequences}
Synthetic sequences are generated by randomly drawing scenarios of recombination from the probability distribution in Eq.~\ref{alphal} or \ref{betah}. In order to fit the data, the resulting sequences are then cut to mimic the sequencing process (e.g. fixed starting point and fixed read length).

\subsubsection{Comparison to other software}
We benchmarked our method against MiXCR 2.0.2 \cite{bolotin2015mixcr} --  a commonly used deterministic alignment method. We used the  MiXCR sequence assignment to compute the frequency of gene usage, insertion length, deletions and obtain the distributions shown in Fig.~\ref{SIFig_compsoft}. We also compared to Partis \cite{ralph_consistency_2016} -- a recent HMM based model of recombination. Since Partis uses a Viterbi learning algorithm, we used the most likely assignments it outputs to compute the corresponding probability distribution shown in Fig.~\ref{SIFig_compsoft}. Since Partis is designed to handle BCRs we assessed its performance on the BCR dataset only.

\begin{figure*}
	\noindent\includegraphics[width=\textwidth]{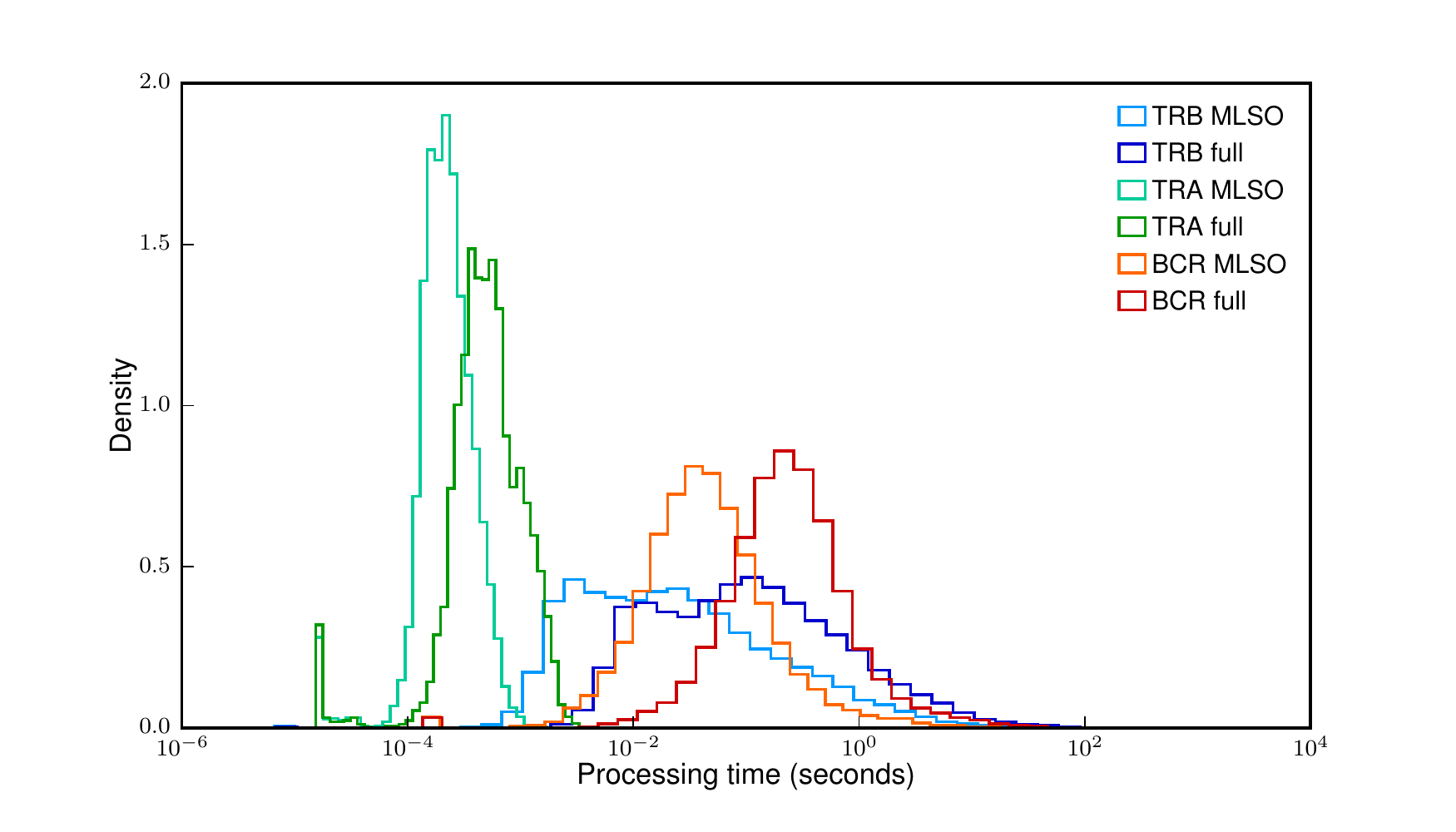}
	\caption{\textbf{Distribution of the processing time per sequence.} Shows the processing time for finding the Most Likely Scenario Only (MLSO) and to evaluate all scenarios (full) for the different chains. Histograms were computed on 20000 sequences for each chain on a single core of an Intel(R) Xeon(R) CPU E5-2680 v3 2.50GHz processor running code compiled with gcc (Debian 4.9.2-10). 
		We benchmarked IGoR's performance for evaluating possible recombination scenarios on real data sequences used to infer the models presented in the main text. We used 60bp TCR $\beta$ sequences for benchmarking since the difficulty for finding the correct V and J for alignment is higher.
		Finding the Most Likely Scenario Only(MLSO) is on average $3\times$ faster than evaluating all possible scenarios. Restricting possible scenarios to deterministically assigned V and J genes is on average $6\times$ faster(data not shown). }
	\label{SIfig_runtime}
\end{figure*}

\begin{figure*}

	\noindent\includegraphics[width=\textwidth]{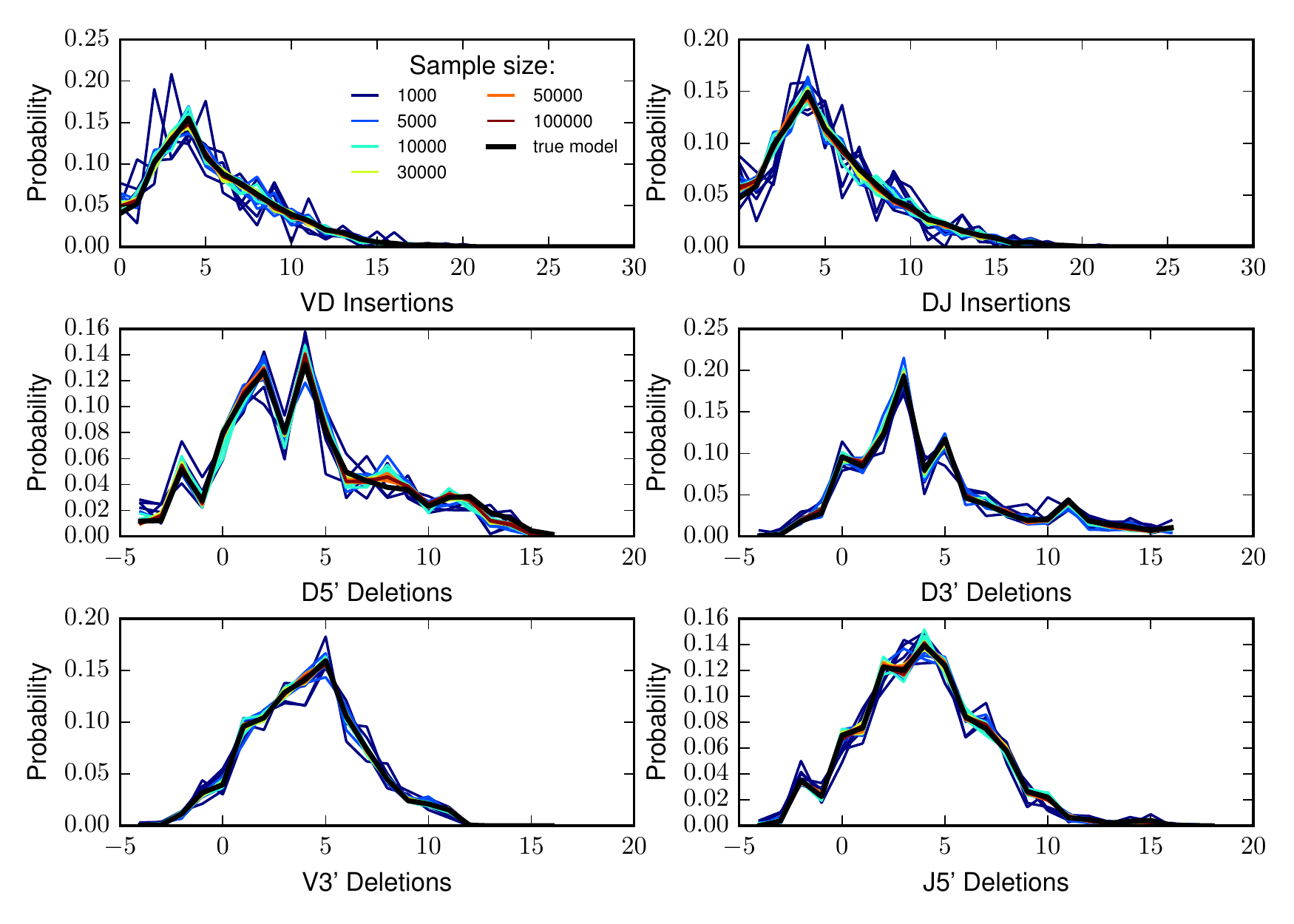}

	\caption{{\bf Tested on simulated data with a known underlying model Igor converges to the true distribution for different error rates. } We show insertion and deletion distributions obtained from 60bp TCR generated samples of various sizes and with various error rates, to underline qualitative differences hidden by the Kullback-Leibler divergence shown in Fig.~\ref{Fig3} and Fig.~\ref{SISampling_DKL_breakup}. 
	}
	\label{SIFig_conv}
\end{figure*}

\begin{figure*}[h]
	\noindent\includegraphics[width=\textwidth]{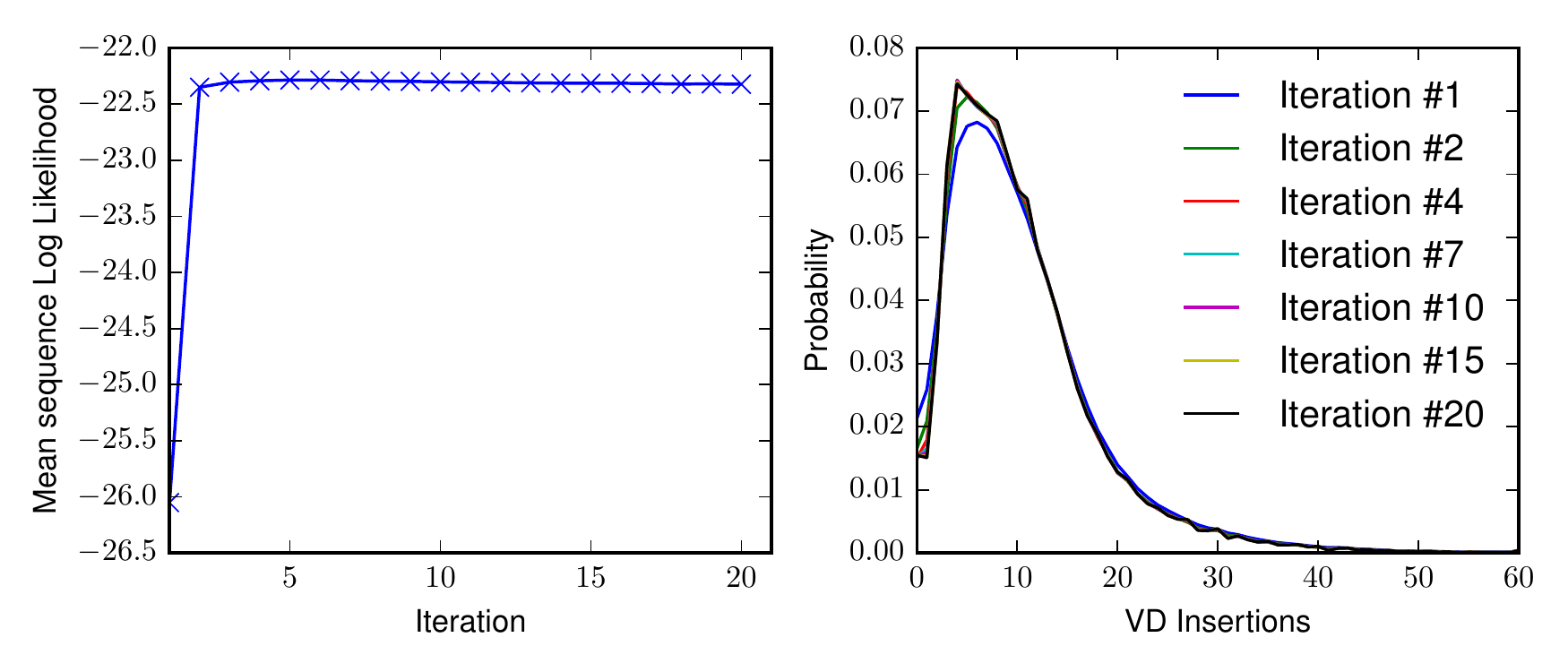}
	\caption{{\textbf{Convergence of IGoR for a naive BCR dataset. }}\textbf{A.} The mean log likelihood per sequence increases and quickly plateau, thus reaching the maximum likelihood estimate of the parameters. \textbf{B.} Convergence of the distribution is shown with the example of the distribution of number of VD insertions.
	\label{SIconvergence_fig}
	}
\end{figure*}

\begin{figure*}[h]
	\noindent\includegraphics[width=\textwidth]{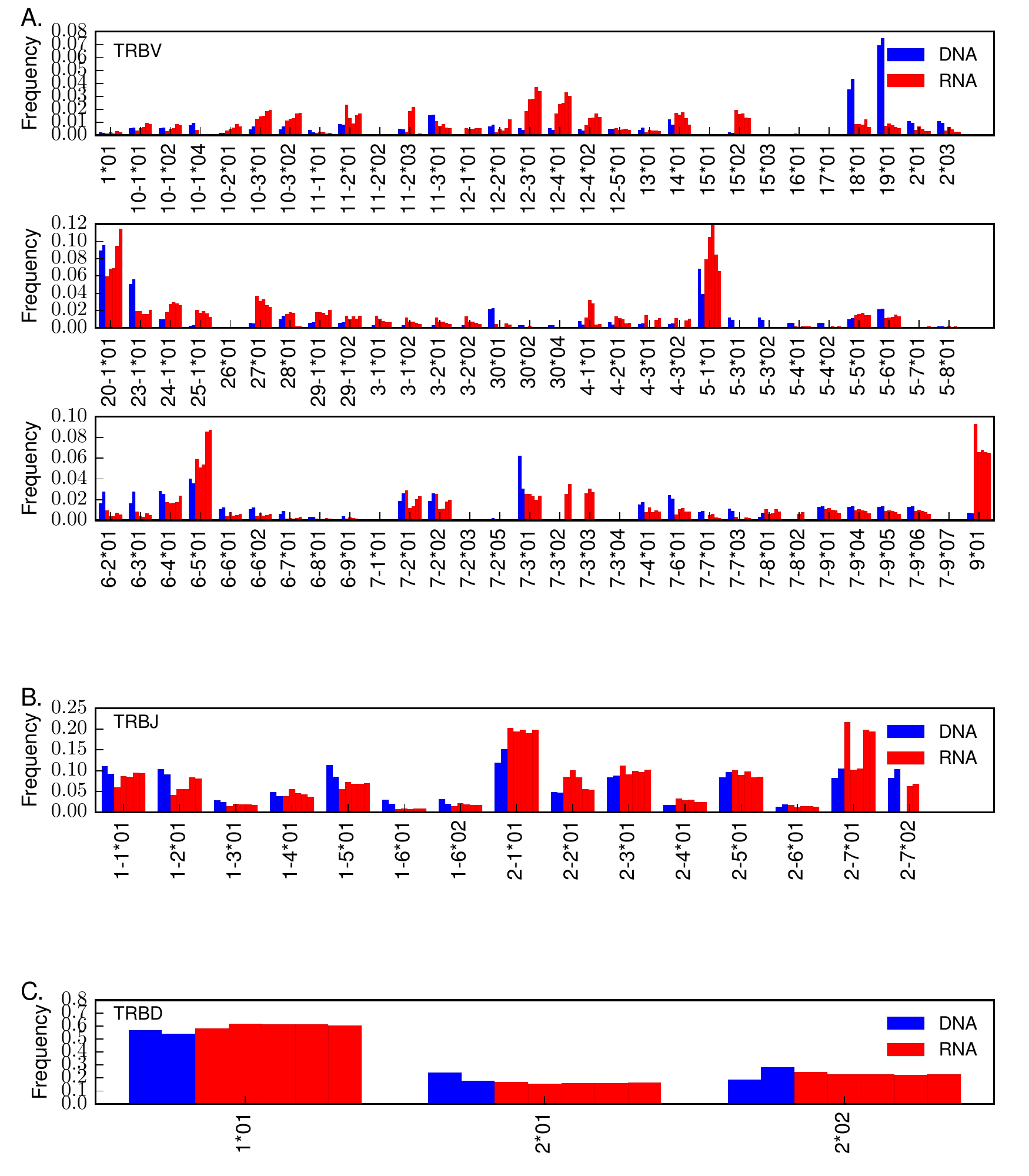}
	\caption{{\bf Gene usage in TRB mRNA vs DNA data.}    We plot the marginal gene usage averaged over conditional dependencies for V, D and J genes respectively inferred using IGoR from mRNA 100bp (red) and DNA 60bp (blue) technology datasets. We observe a higher inter-method than inter-individual variability. 
		\label{SIFig_TCRmRNADNA}
	}
\end{figure*}

\begin{figure*}
	\noindent\includegraphics[width=\textwidth]{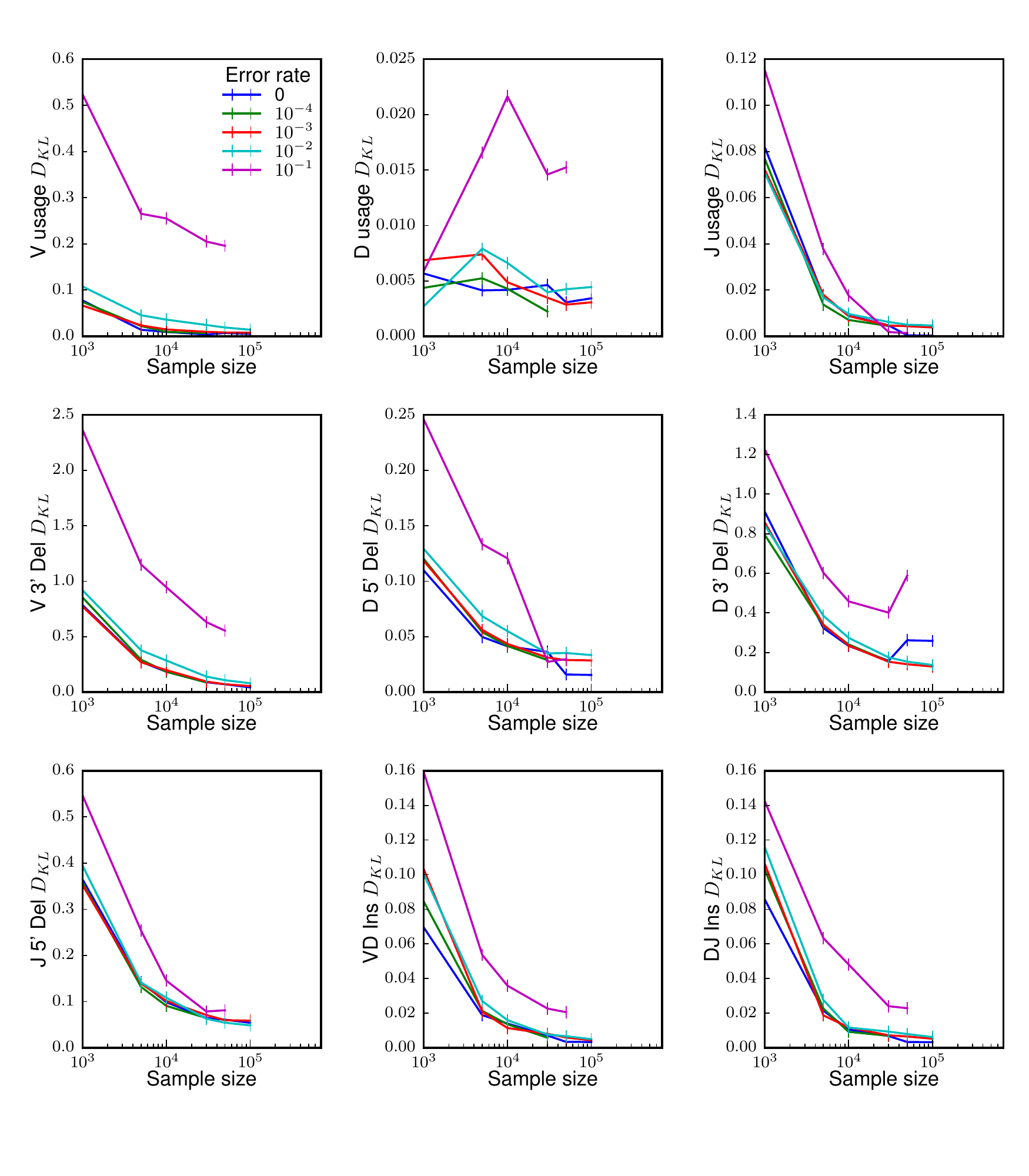}
	\caption{{\bf Synthetic sampling $D_{KL}$ breakup} Kullback-Leibler divergence ($D_{\text{KL}}(\rm{inferred}\ ||\ \rm{true})$) in bits between models inferred on various sample sizes of sequences with various error rates and the ground truth decomposed for the different model components. All components reach a small divergence value for sufficiently large sample sizes.}    
	\label{SISampling_DKL_breakup}
\end{figure*}

\begin{figure*}[h]
	\noindent\includegraphics[width=\textwidth]{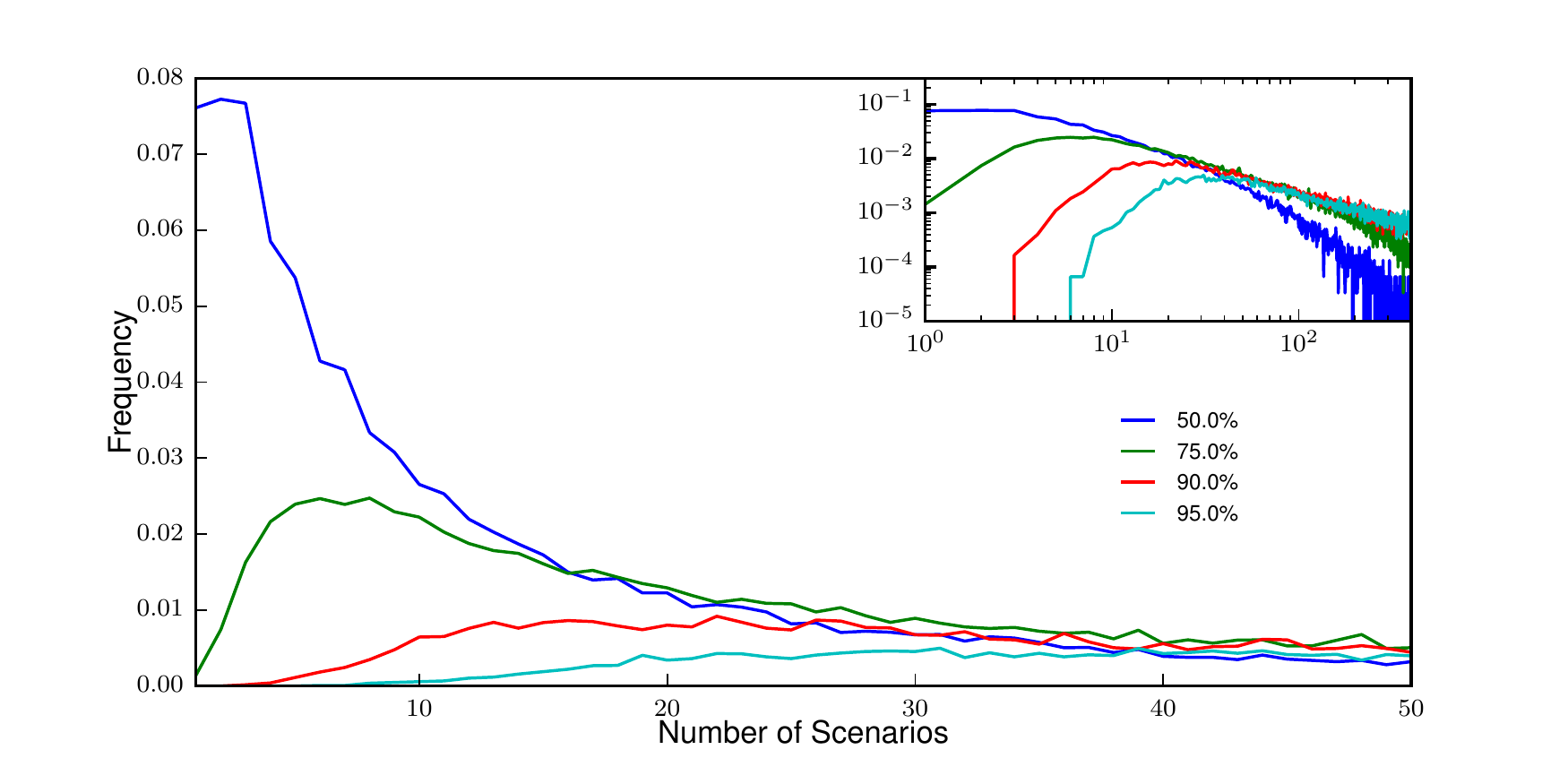}
	\caption{{\bf A probabilistic assignment approach is crucial for TCRs. } Equivalent of main text Fig.~\ref{Fig4}b for 30000 60bp TCRs. This figure shows the distribution of the number of scenarios that need to be enumerated (from most to least likely) to include the true scenario with 50\% (blue), 75\% (green), 90\% (red), or 95\% (cyan) confidence. The shorter read length compared to 130bp BCRs entail a higher uncertainty on the V gene identity, for which a higher number of scenarios must be considered. 
	}
 \label{SIFig_probasTCR}
\end{figure*}

\begin{figure*}[h]
	\noindent\includegraphics[width=\textwidth]{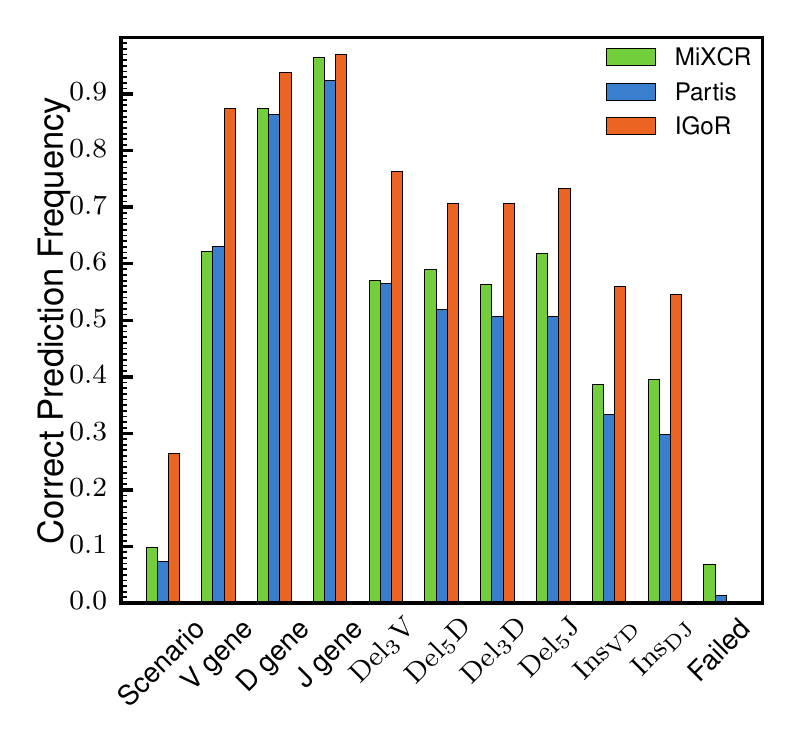}
	\caption{{\bf Assignment performance on sequences without palindromic insertions} We have shown in main text Fig.~\ref{Fig4}c the ability of MiXCR, Partis and IGoR to predict the correct scenario of recombination. Since Partis does not model palindromic insertions we here present the performance of the three software one sequences that were generated without any. Although Partis' prediction is improved and reaches 7.4\% close to MiXCR's 9.8\% accuracy, both remain lower than IGoR's 26.5\% correct predictions.  
		\label{SIFig_no_Pnuc_assign.pdf}
	}
\end{figure*}

\begin{figure*}
	\noindent\includegraphics[width=\textwidth]{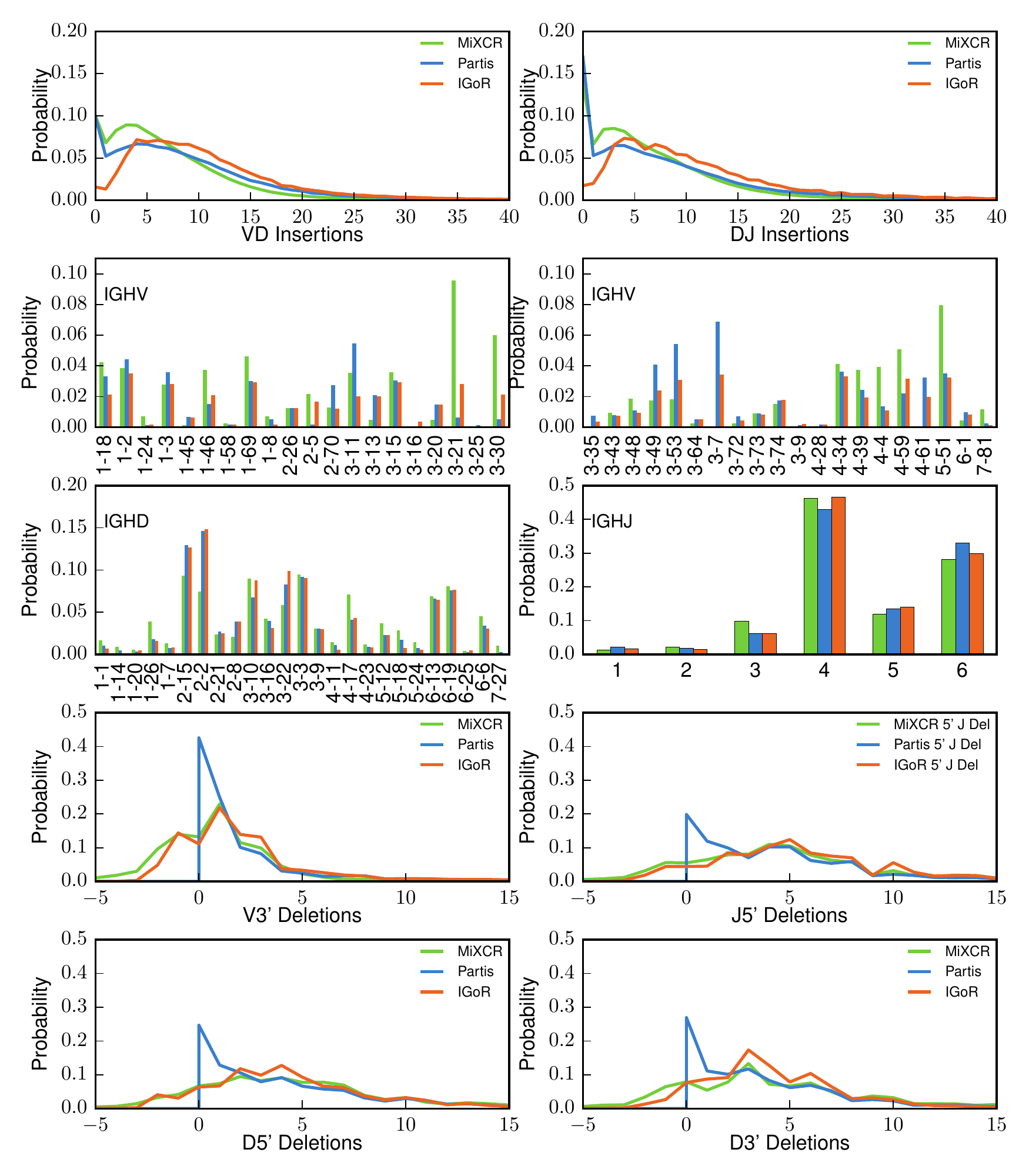}
	\caption{{\bf Comparison of distributions obtained from different softwares.} MiXCR performing deterministic alignments and Partis Viterbi learning we used the output assignments to compute the corresponding recombination statistics. We plot them along with IGoR's distribution obtained from our maximum likelihood approach. Note that for ease of presentation we show distributions averaged over conditional dependences. From the two top panels we observe that Partis and MiXCR overestimate the frequency of low number of non templated insertions. Gene usage is mostly consistent between methods. In the four bottom panels, negative number of deletions denote palindromic insertions. We observe that the three methods obtain qualitatively different marginal distribution of number of deletions.  
	}
	\label{SIFig_compsoft}
\end{figure*}

\begin{figure*}[h]
	\noindent\includegraphics[width=\textwidth]{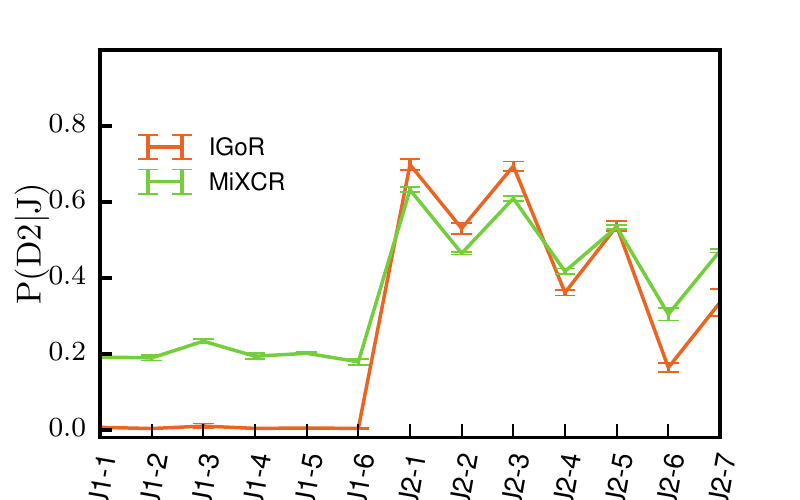}
	\caption{\textbf{Data TRB D2-J association.} As we have shown the D,J pairing rule for TCRs on synthetic data in main text Fig.~\ref{Fig4}D, we plot here the distributions $P(D2|J)$ obtained on real 100bp TCR mRNA data for IGoR and MiXCR. Again, IGoR is able to capture the physiological exclusion between D2 and J1 while MiXCR is not.}
	\label{SIFig_TCR_DJ}
\end{figure*}

\begin{figure*}[h]
	\noindent\includegraphics[width=\textwidth]{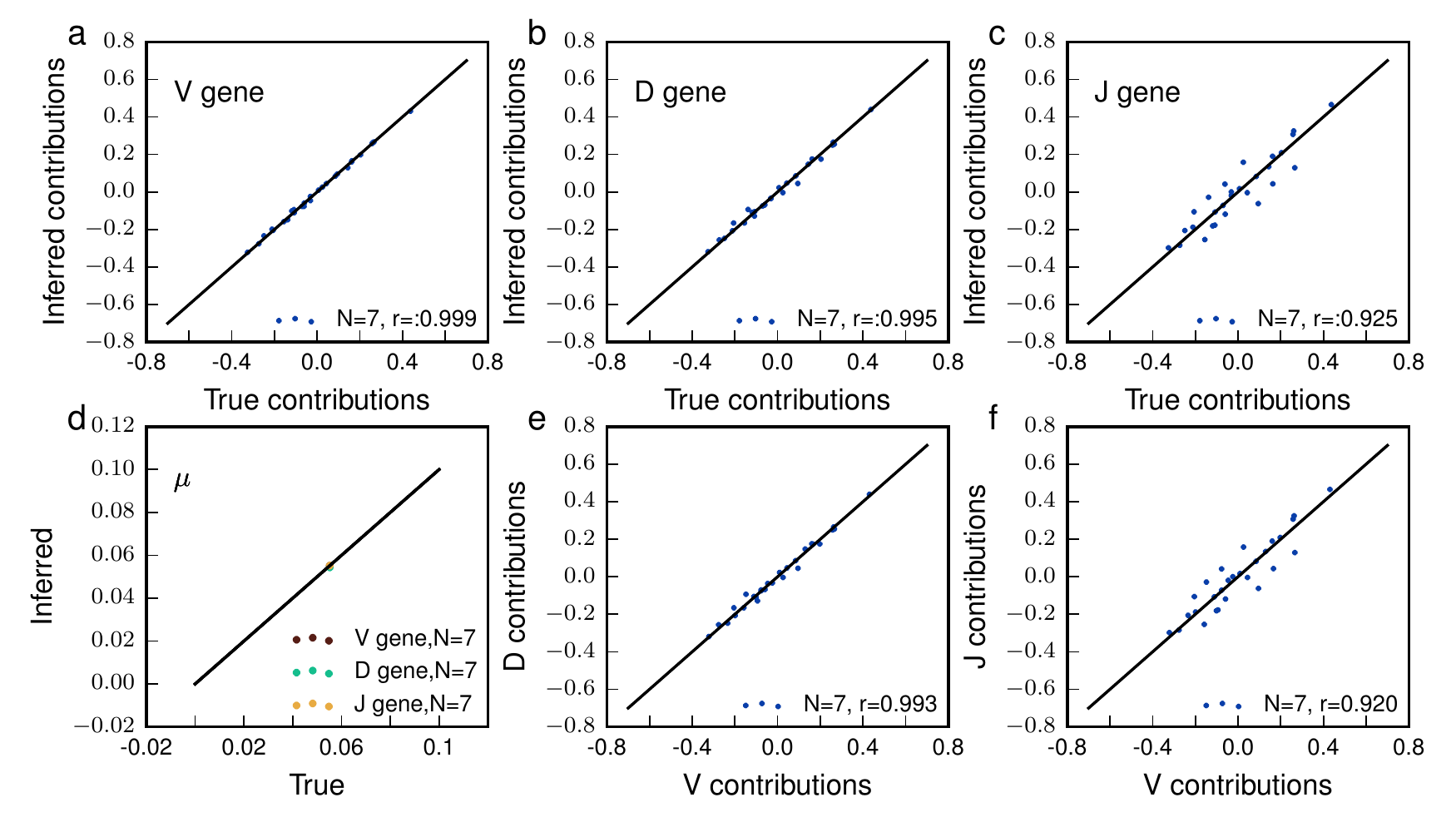}
	\caption{\textbf{Inference of the 7mer hypermutation model on synthetic sequences.} In order to assess the validity of our method we generate synthetic BCRs sequences from a heavy chain model learned on naive data sequences. We then generate Poisson distributed errors on the sequences by simulating mutations at each base pair by a Bernouilli process according to the hypermutation model learned on the V gene of memory sequences. We then cut the sequences in 130bp reads in order to mimic real data sequences. The results of this experiment shows that the model can be perfectly inferred on V and D genes while the scatter on J gene is higher. This can be explained by the limited number of n-mers observed on J gene since sequences were cut to mimic sequencing from a primer in the J. }
	\label{SIFig_hyperm_synth}
\end{figure*}

\begin{figure*}[h]
	\noindent\includegraphics[width=\textwidth]{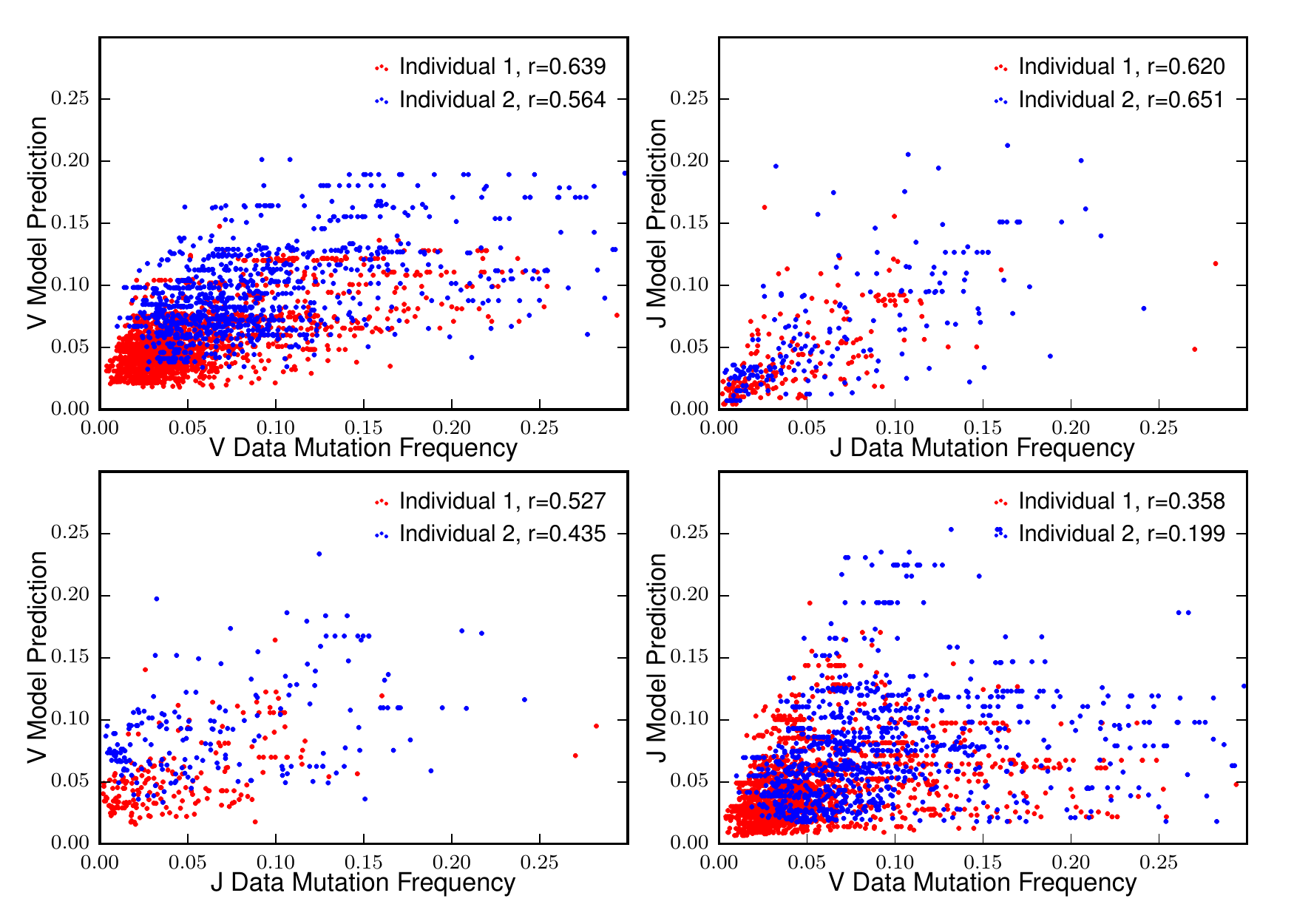}
	\caption{
		\textbf{Prediction of the mutation frequencies on real data.}
		By direct exploration of recombination scenarios we recorded the posterior mutation frequency per individual base pairs on V and J genomic templates and compare it to the independent 7-mer model. We plot a scatter for base pairs that have been observed at least 2000 times on a 100 000 sequences dataset, for which we can compute a reliable mutation frequency, and the mutation frequency predicted by our model. The two top panels show good predictive power for the gene on which the model was learned. However the two bottom panels show a lesser ability to predict the correct mutation frequencies on the whole locus, hence suggesting that differences observed in inferred position weight matrices (Fig.~\ref{SIFig_all_logos}) are of biological relevance. 
		}
	\label{SIFig_mut_freq_pred}
\end{figure*}

\begin{figure*}[h]
	\noindent\includegraphics[width=\textwidth]{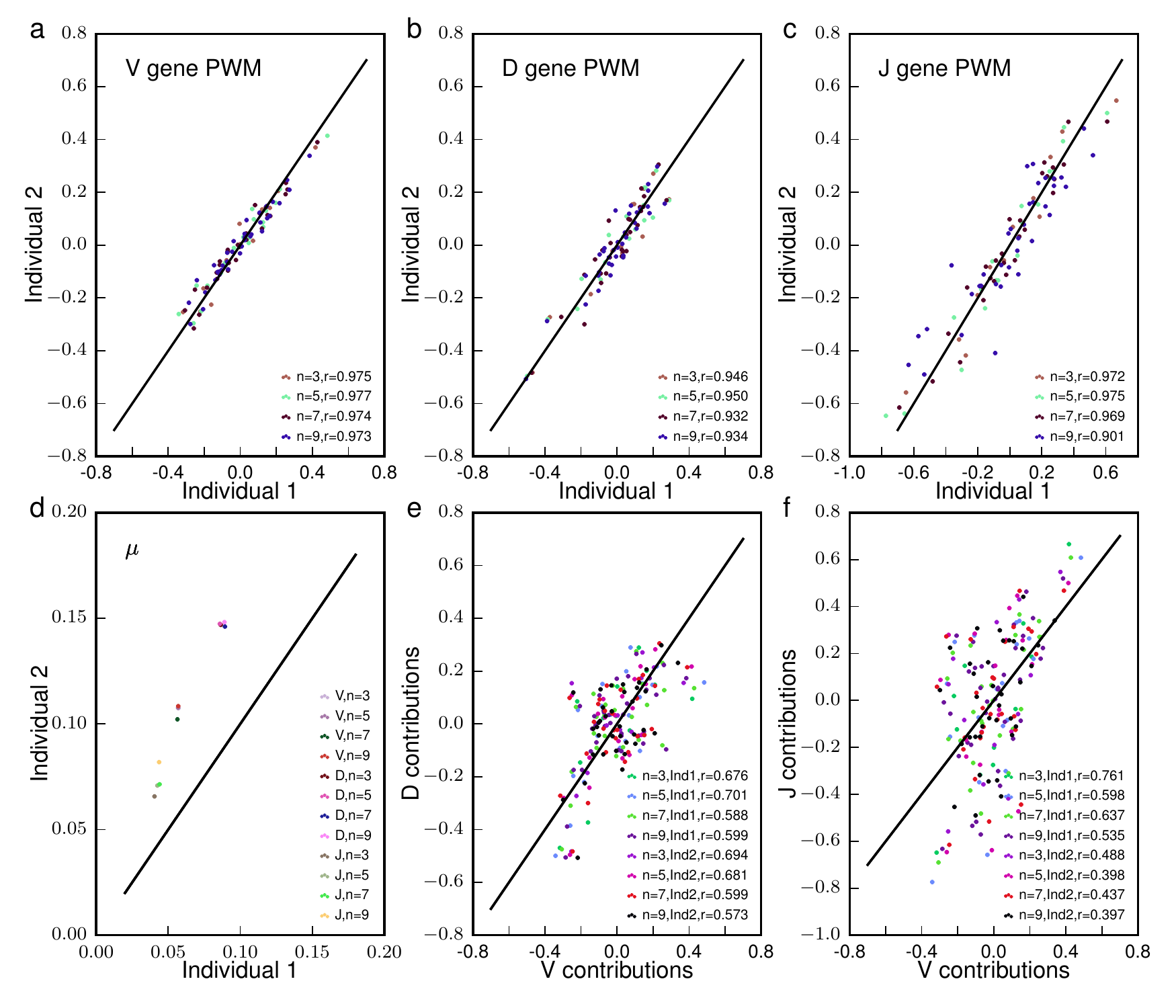}
	\caption{\textbf{Inference of the hypermutation model on real non productive memory BCR sequences.} 
		\textbf{a}, \textbf{b} and \textbf{c} compare the position weight matrices inferred on respectively V, D and J genes for different n-mer length. For all sizes and gene the inferred contributions are extremely reproducible from an individual to the other.
		\textbf{d} Comparison of the overall mutational load on different individuals and gene for different n-mer size. This overall mutational load varies from individual to individual and within the locus.
		\textbf{e} and \textbf{f} Comparison between contributions inferred on different genes. We observe weaker inter gene correlations than the one observed for inter individual contributions. 
		 }
	\label{SIFig_hyperm_real}
\end{figure*}

\begin{figure*}
	\noindent\includegraphics[width=\textwidth]{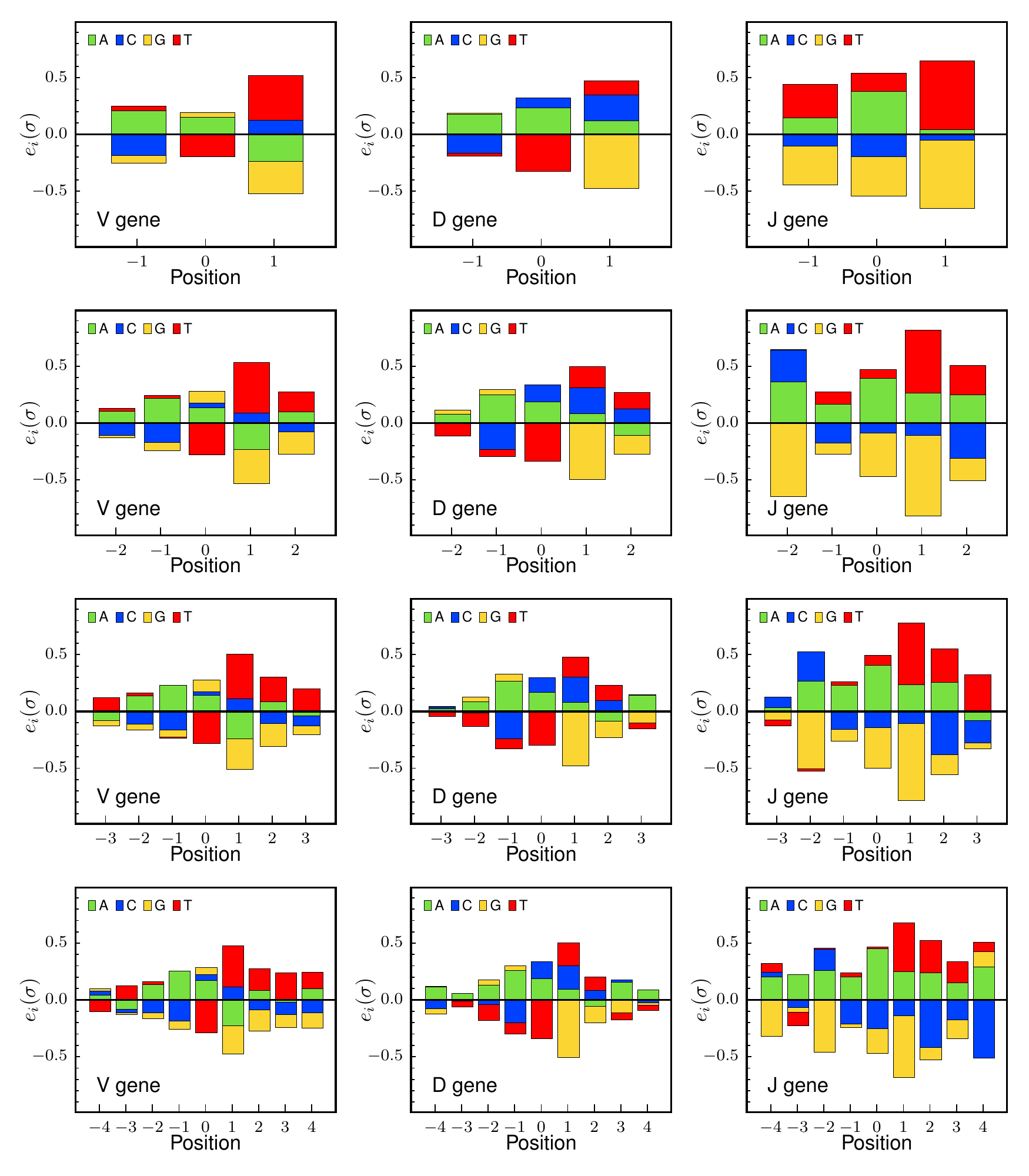}
	\caption{
		\textbf{Context logo for different context sizes on the three different genes.}
		We inferred position weight matrices for different n-mer sizes for V, D and J. With increasing n-mer sizes, side contributions do not vanish. 
		}
	\label{SIFig_all_logos}
\end{figure*}

\begin{figure*}[h]
	\noindent\includegraphics[width=\textwidth]{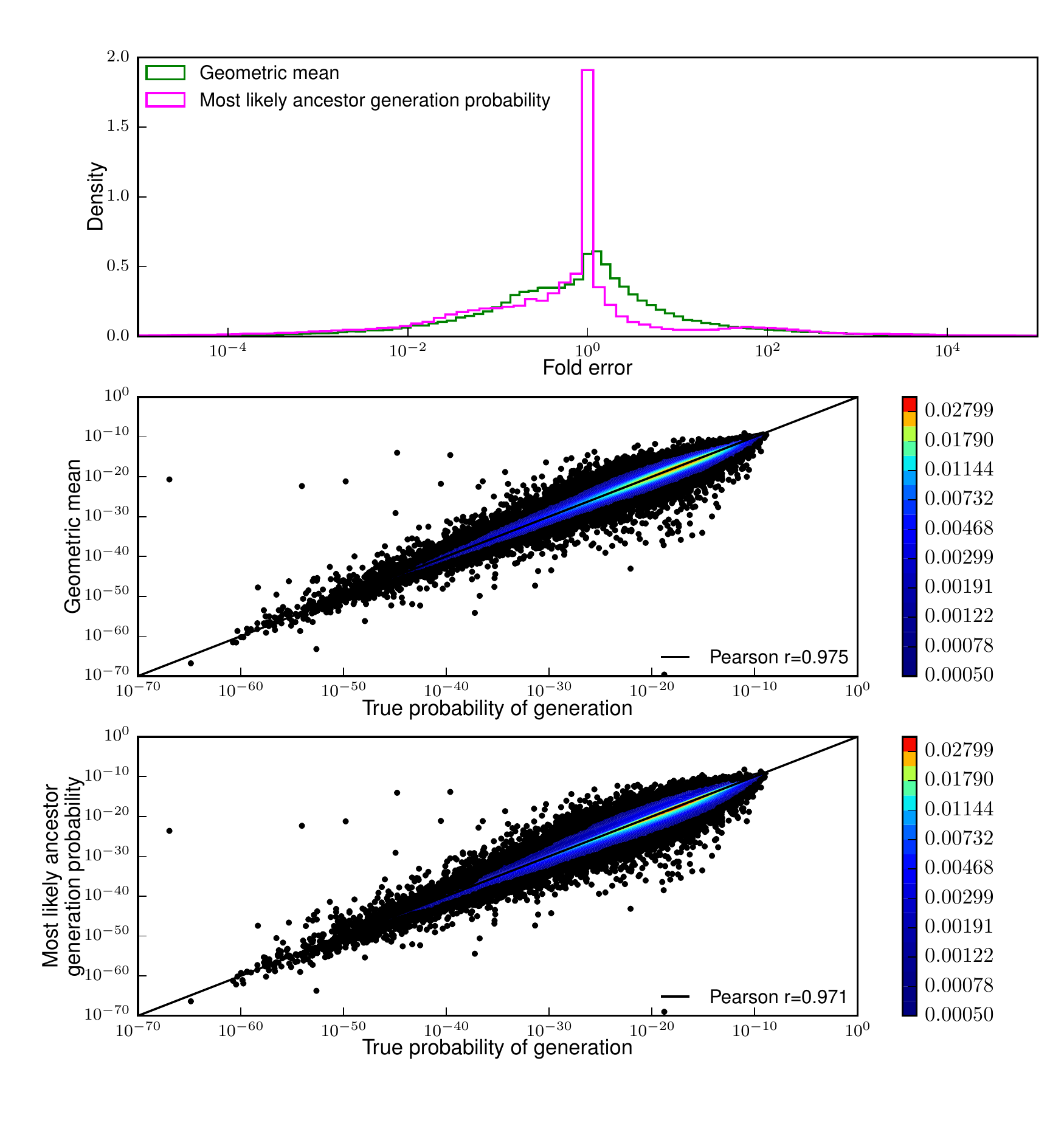}
	\caption{{\bf Sequence probability of generation estimation}  
		By generating synthetic 130bp BCR sequences from an inferred recombination model without errors we were able to compute their probability of generation $P_{gen}$ (see SI \ref{perfect_seq}). We further introduced errors in those sequences, errors whose statistics correspond to an inferred hypermutation model and computed an estimate for the probability of generation of the unmutated ancestor. We propose two different estimators: $\overline{P_{gen}}$ a geometric average of putative ancestors probability of generation weighted by it's posterior probability (green and middle) and $P_{gen}(\underset{\vec{S}}{argmax}P(\vec{S}|\vec{r}))$ the probability of generation of the most likely ancestor (pink and bottom). Note that due to convergent recombination the most likely ancestor does not necessarily correspond to the sequence implied by the most likely scenario. Thus these two estimates can only be made thanks to direct exploration of recombination scenarios. Both estimators show almost perfect correlation despite the error distribution of most likely ancestor probability of generation being non symmetric. 
	}
	\label{SIFig_Pgen}
\end{figure*}

\begin{figure*}[h]

	\noindent\includegraphics[width=\textwidth]{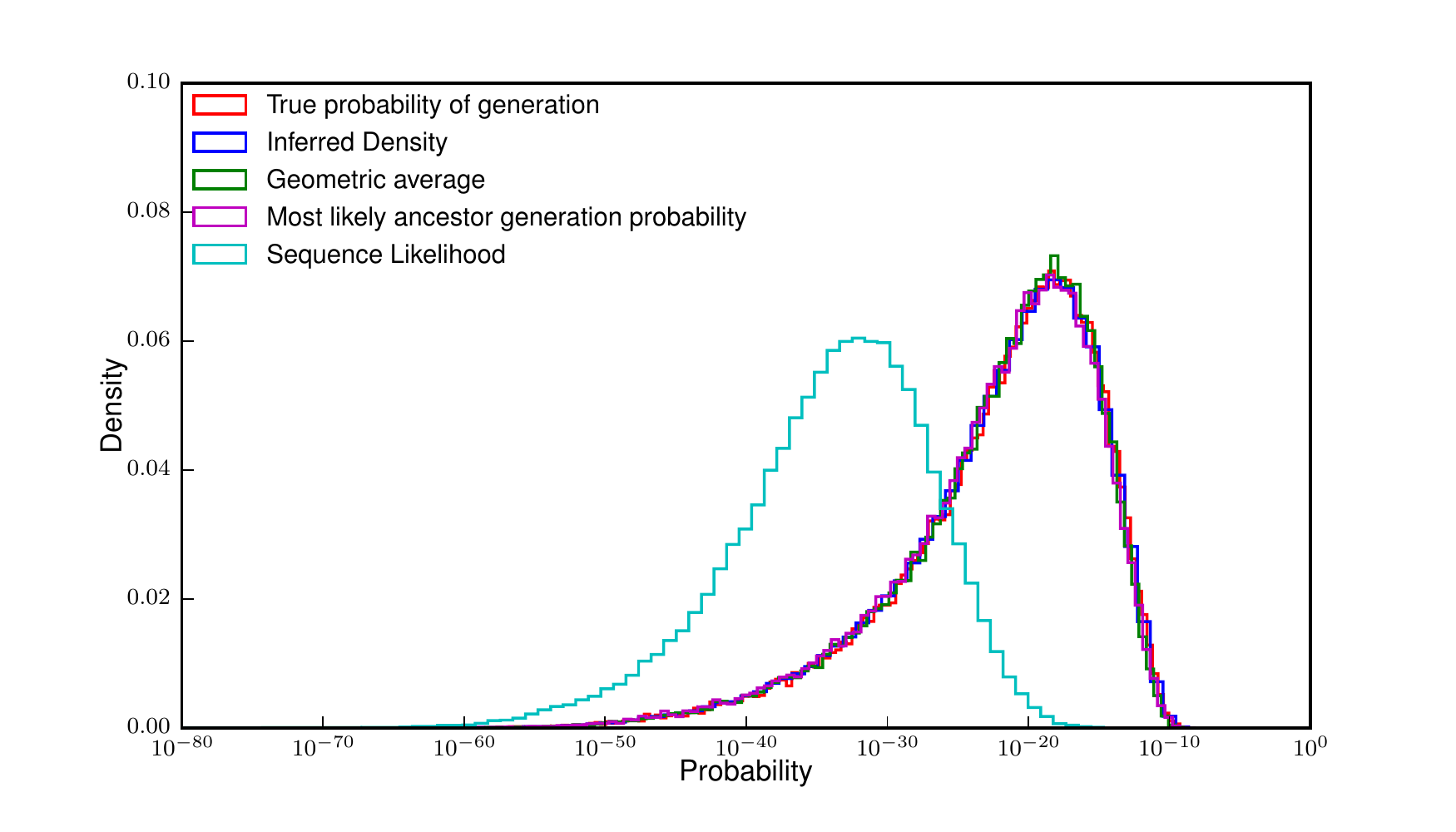}
	\caption{\textbf{ Density of the probability of generation of sequences}  
		We plot the distribution of probability of generation obtained from different estimators against the true distribution of generation probabilities. The true probability of generation, the geometric average and the probability of generation of the most likely ancestor are presented in Fig.~\ref{SIFig_Pgen}'s caption. The inferred density (blue) is a histogram of each sequence putative ancestors probability of generation weighted by it's posterior probability. We also plot the distribution of sequence likelihoods, that could be obtained by other methods (e.g forward algorithm) and show that it greatly differs from the distribution of generation probability. 
	}
	\label{SIFig_Pgen_density}
\end{figure*}

\begin{figure*}
	\noindent\includegraphics[width=\textwidth]{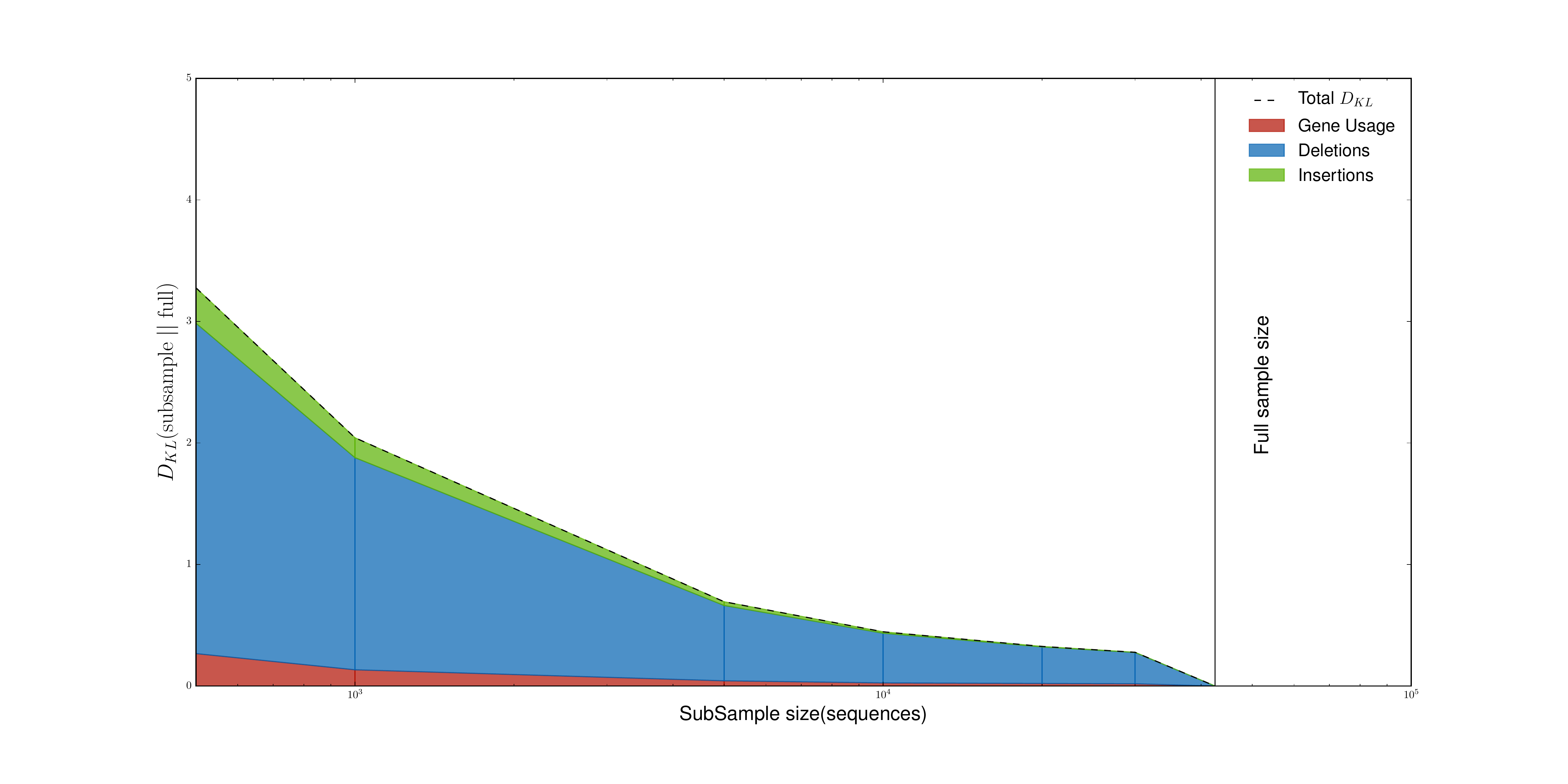}
	\caption{{\bf Bootstrap} Kullback-Leibler divergence ($D_{\text{KL}}(\rm{subsample}\ ||\ \rm{full})$) in bits between the model inferred on the full data sample and models inferred on various subsamples sizes.}    
	\label{SIFig_sampling}
\end{figure*}

\begin{figure*}[h]
	\noindent\includegraphics[width=\textwidth]{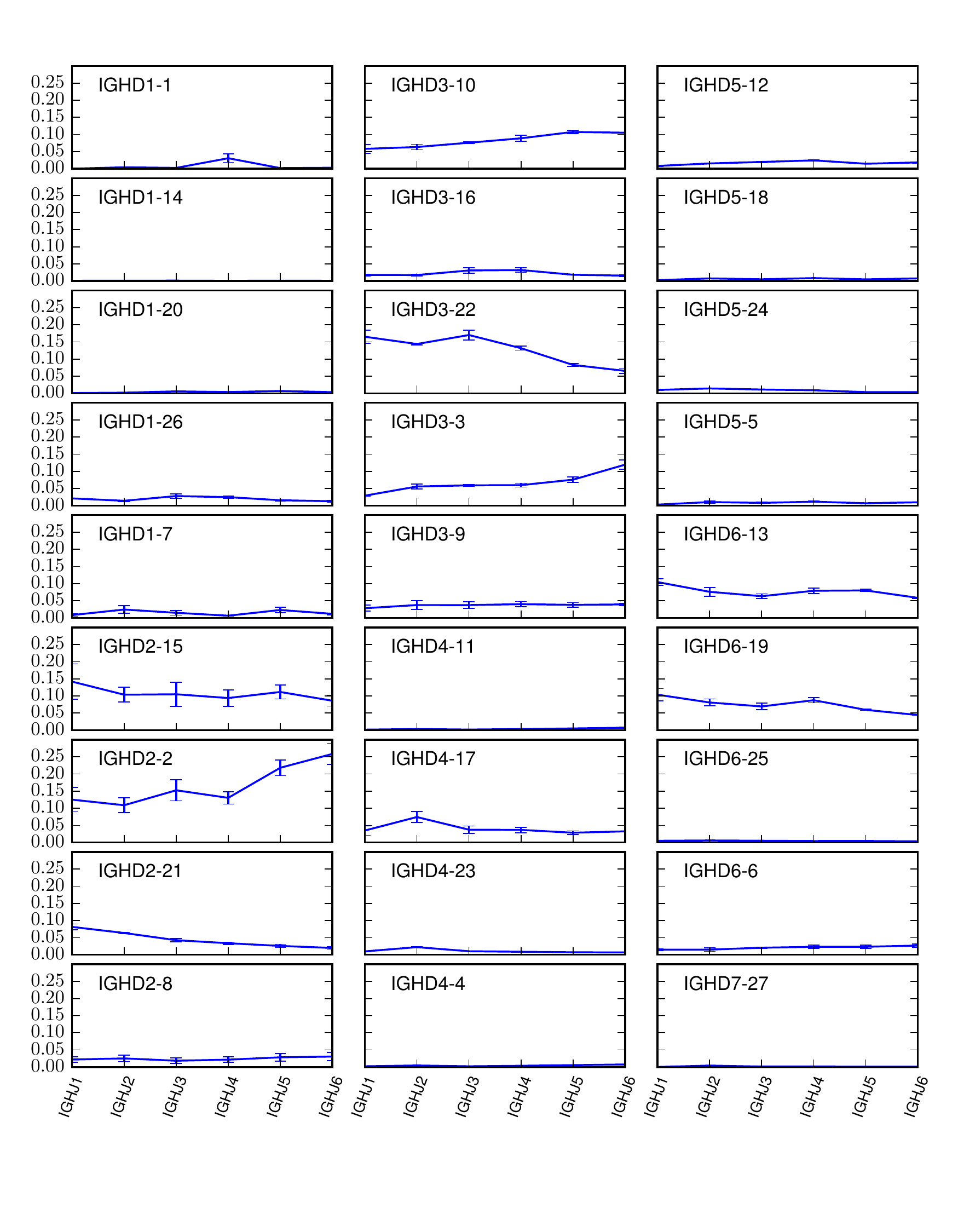}
	\caption{\textbf{BCR D-J association.} As we have shown the D,J pairing rule for TCRs in main text Fig.~\ref{Fig4}d, we plot $P(D|J)$ for each pair. Unlike TCRs, BCRs do not seem to exhibit such a clear coupling.}
	\label{SIFig_BCR_DJ}
\end{figure*}

\begin{figure*}[h]
	\noindent\includegraphics[width=\textwidth]{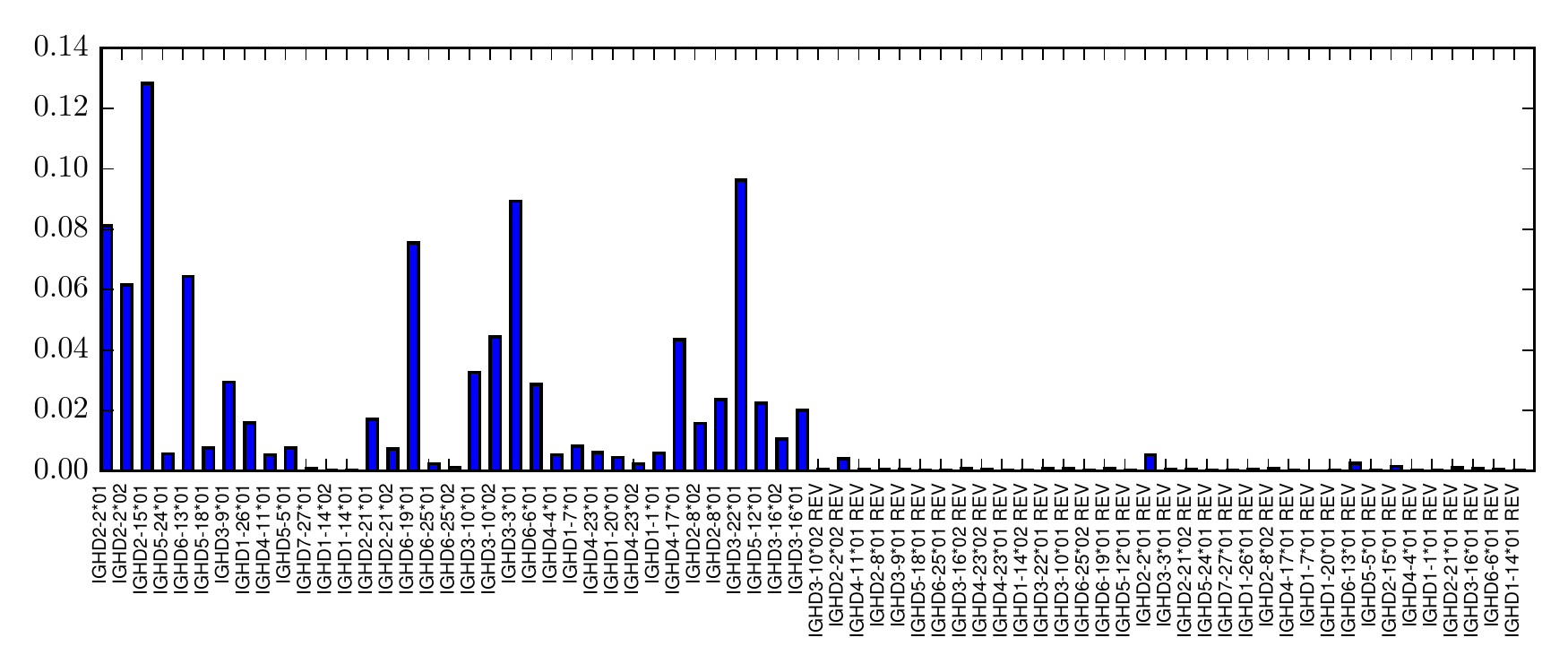}
	\caption{ \textbf{BCR reversed complement Ds usage.} By appending the reversed complement of each D gene to the list of D genes we have tested the occurrence of reversed Ds during the VDJ recombination process. We can see that although some reversed complement Ds can be observed the effect is minor.}
	\label{SIFig_revDs}
\end{figure*}

\end{document}